\documentclass[journal]{IEEEtran}

\usepackage[utf8]{inputenc} 
\usepackage[T1]{fontenc}    
\usepackage{hyperref}       
\usepackage{url}            
\usepackage{booktabs}       
\usepackage{amsfonts}       
\usepackage{nicefrac}       
\usepackage{microtype}      
\usepackage{color}
\usepackage{algorithm,algpseudocode}
\usepackage{subcaption}
\usepackage{graphicx}
\usepackage{amsmath}
\usepackage{float}
\usepackage{stfloats}
\usepackage{cite}
\usepackage{multirow}
\usepackage{amssymb}
\usepackage{amssymb}
\usepackage{pifont}
\newcommand{\cmark}{\ding{51}}%
\newcommand{\xmark}{\ding{55}}%
\usepackage{color}
\usepackage{cite}
\usepackage[hang,flushmargin]{footmisc}

\ifCLASSINFOpdf
\else
\fi
\begin{document}
%
\title{Audio-visual Multi-channel Integration and Recognition of Overlapped Speech}
%
%
%

\author{Jianwei Yu,~\IEEEmembership{}
        Shi-Xiong Zhang,~\IEEEmembership{Member,~IEEE,}
        Bo Wu,~\IEEEmembership{}
        Shansong Liu,~\IEEEmembership{}
        Shoukang Hu,~\IEEEmembership{}
        Mengzhe Geng,~\IEEEmembership{}\\
        Xunying Liu,~\IEEEmembership{Member,~IEEE,}
        Helen Meng,~\IEEEmembership{Fellow,~IEEE}
        Dong Yu,~\IEEEmembership{Fellow,~IEEE}
\thanks{
Half of the work leading to this research was conducted while Jianwei Yu was an intern at Tencent Lab, Shenzhen, China.  He is now a Ph.D student in the Chinese University of Hong Kong, Hong Kong (email: jwyu@se.cuhk.edu.hk). Professor Xunying Liu is the corresponding author.
}
\thanks{
Shansong Liu, Shoukang Hu, Mengzhe Geng, Xunying Liu and Helen Meng are with the Department of System Engineering and Engineering Management. 
The Chinese University of Hong Kong, Hong Kong (email: xyliu@se.cuhk.edu.hk; hmmeng@se.cuhk.edu.hk). 
}
\thanks{Shi-Xiong Zhang, Bo Wu and Dong Yu are with the Tencent AILab  (email: \{auszhang, lambowu, dyu@tencent.com)}
}

%
%

\markboth{Journal of \LaTeX\ Class Files,~Vol.~14, No.~8, August~2015}%
{Shell \MakeLowercase{\textit{et al.}}: Bare Demo of IEEEtran.cls for IEEE Journals}
%



\maketitle

\begin{abstract}
Automatic speech recognition (ASR) technologies have been significantly advanced in the past few decades. However, recognition of overlapped speech remains a highly challenging task to date.
To this end, multi-channel microphone array data are widely used in current ASR systems. 
Motivated by the invariance of visual modality to acoustic signal corruption and the additional cues they provide to separate the target speaker from the interfering sound sources, this paper presents an audio-visual multi-channel based recognition system for overlapped speech. 
It benefits from a tight integration between a speech separation front-end and recognition back-end, both of which incorporate additional video input.
A series of audio-visual multi-channel speech separation front-end components based on \textit{TF masking},  \textit{Filter\&Sum} and \textit{mask-based MVDR} neural channel integration approaches are developed. 
To reduce the error cost mismatch between the separation and the  recognition components, the entire system is jointly fine-tuned using a multi-task criterion interpolation of the scale-invariant signal to noise ratio (Si-SNR) with either the connectionist temporal classification (CTC), or lattice-free maximum mutual information (LF-MMI) loss function.
Experiments suggest that: the proposed audio-visual multi-channel recognition system outperforms the baseline audio-only multi-channel ASR system by up to 8.04\% (31.68\% relative) and 22.86\% (58.51\% relative) absolute WER reduction on overlapped speech constructed using either simulation or replaying of the LRS2 dataset respectively.
Consistent performance improvements are also obtained using the proposed audio-visual multi-channel recognition system when using occluded video input with the lip region randomly covered up to 60\%.

\end{abstract}


%
\IEEEpeerreviewmaketitle

\section{Introduction}
%
%
%
%

\IEEEPARstart{D}{espite} the rapid progress of automatic speech recognition (ASR) technologies in the past few decades, recognition of overlapped speech remains a highly challenging task.
The presence of interfering speakers creates a large mismatch between the target speaker’s clean speech and the mixed signal. This often leads to large performance degradation of current ASR systems. 
To this end, acoustic beamforming techniques integrating sensor data from multiple array channels are widely used.
These multi-channel array signal integration approaches are normally implemented as time or frequency domain linear filters that are capable of “listening” in the target speaker’s direction while minimizing the effects of noise distortions and other interfering speakers from other directions.
The desired target speech signal is thus enhanced.

Microphone arrays play a key role in state-of-the-art ASR systems designed for multi-talker overlapped and far field speech \cite{CHIME3, NTTchime3, harper2015automatic, ROM, MIMO2019, MIMO2020}, often following a traditional speech enhancement prior to recognition based system architecture.
These systems contain two separately developed components: the speech separation and enhancement front-end module, and the speech recognition back-end. These two components are often integrated in a pipelined manner.
The separation front-end module is often implemented using conventional beamforming techniques represented by either time domain delay and sum \cite{DSB, beamformit} or frequency domain minimum variance distortionless response (MVDR) \cite{MVDR2001, MVDR} and the related generalized eigenvalue (GEV) \cite{GEV2007} channel integration approaches. 
The former uses generalized phase correlation between sensor inputs and a Viterbi search procedure to estimate the optimal channel delays and their respective combination weights.
The frequency domain beamforming approaches maximizes the signal to noise ratio (SNR) of the filtered outputs. 

With the successful and wider application of deep learning based speech technologies, microphone array channel integration methods have evolved into a variety of neural network (NN) based designs in the past few years.
These NN based methods can be classified into three main categories including \textit{TF masking}, \textit{Filter\&Sum} and \textit{mask-based MVDR} or \textit{GEV}.
In contrast to the traditional mask based single channel speech separation methods \cite{WANG2018, PIT2017}, multi-channel information is fed into DNNs in the TF masking approaches \cite{WJTVF, MULTIBAND} to predict spectral time-frequency (TF) mask labels for a reference channel that specify whether a particular TF spectrum point is dominated by the target speaker or interfering sources to facilitate speech separation.
The neural \textit{Filter\&Sum} approaches directly estimate the beamforming filter parameters in either time domain \cite{Taramulti, FASNET1, FASNET2} or frequency domain \cite{DEEPBEAMFORMER} to produce the separated outputs. 
The \textit{mask-based MVDR} \cite{wang2018spatial, IMPMVDR, Angle, ROM, MULTIFAROVER, MIMO2019, MIMO2020} and related \textit{mask-based GEV} \cite{GEV2016, Beamnet} approaches predict the TF masks using DNNs before estimating the power spectral density (PSD) matrices for the target and overlapping speakers to obtain the beamforming filter parameters.
Compared with the conventional stand-alone beamforming approaches, these neural based methods allow a tighter integration with the downstream recognition back-end \cite{DEEPBEAMFORMER, Beamnet, CRM, MIMO2019, MIMO2020}.
Large performance improvements have been reported for overlapped speech recognition tasks by using microphone array based multi-channel inputs \cite{MIMO2019, MIMO2020}. 
However, the current systems’ performance gap between overlapped and non-overlapped speech remains large.

Inspired by the bi-modal nature of human speech perception, there has been increasing interest in incorporating visual information into the speech separation and recognition systems for far-field and overlapped speech. 
The advantages of these approaches are three folds: 
1) visual information contains additional cues such as lip movements that differentiate the target speech from other interfering sources; 
2) lip movements can provide further information over articulation to improve phonetic discrimination; 
3) visual modality is usually invariant to the acoustic signal corruption in noisy or multi-talker environment.
Previous research has successfully used visual information to improve single-channel overlapped speech separation \cite{AVSE1, AVSEWJ, MYLIP} and recognition \cite{AVSR2016OVERLAP, AVSRJIAN, AVSR2020HAN} performance.
Recently there has also been increasing efforts in developing audio-visual multi-channel input based speech enhancement systems designed for speech separation \cite{gu2020multi} and de-reverberation \cite{tan2019AV}. 
However, currently there is a lack of holistic, full incorporation of visual information into both the front-end speech separation module and the back-end recognition component. 


Performance of audio-visual overlapped speech separation and recognition systems crucially depends on the quality of the video input in terms of the complementary information being provided on top of the audio.
Such sensitivity can be demonstrated by, for example, when the mouth area is obstructed by a mask (often required in the current pandemic), a microphone, or if the speaker stands far away from the camera (low-resolution video inputs).
Only limited previous research was conducted to investigate the system fragility to the aforementioned video occlusion and low-resolution video input problems.
In \cite{DFMSN2019AV}, the authors applied dropout to parts of the DNN acoustic model of a single channel audio-visual speech recognition system connected to the video input during model training to improve the resulting system’s robustness.
In \cite{MYLIP}, the comparable effect in audio-visual speech separation was investigated by using video with artificial occlusion over the mouth region. 
However, there has been very limited previous research investigating the performance sensitivity to video occlusion and low-resolution video inputs in the context of a complete audio-visual multi-channel recognition system of overlapped speech.


In order to address the above issues, an audio-visual multi-channel overlapped speech recognition system featuring tightly integrated separation front-end and recognition back-end is proposed in this paper.
Firstly, for the speech separation front-end, a series of audio-visual microphone array channel integration methods including \textit{TF masking}, \textit{Filter\&Sum} and \textit{mask-based MVDR} are proposed respectively.
Secondly, in order to reduce the error cost mismatch between the separation and the  recognition components that are traditionally trained on different objective functions, they are jointly fine-tuned using a multi-task criterion interpolation of the scale-invariant signal to noise ratio (Si-SNR) with either the connectionist temporal classification (CTC) \cite{CTC}, or lattice-free maximum mutual information (LF-MMI) \cite{LFMMI, pychain} loss function. 
Thirdly, this paper investigates the influence of visual occlusion and low-resolution visual inputs on the proposed systems.
To improve the robustness of audio-visual multi-channel speech recognition systems to visual occlusion, both angle features provided by video cameras mounted on a microphone array and multi-style training consisting of occluded video of lip region coverage up to 80\% are used. In addition, the image in-painting technique \cite{inpainting} is also investigated to restore the occluded video inputs for the visual occlusion issue.  
Experiments suggest that: 
1) the proposed audio-visual multi-channel recognition system outperforms the baseline audio-only multi-channel ASR systems by up to 8.04\% (31.68\% relative) and 22.86\% (58.51\% relative) absolute WER reduction on overlapped speech constructed using either simulation or replaying of the LRS2 dataset; 
2) consistent performances improvements are obtained across all audio-visual multi-channel systems when multi-task criterion based joint fine-tuning is used in place of a  pipelined configuration.
In particular the jointly fine-tuned audio-visual multi-channel system using mask-based MVDR beamforming produced WER reductions by up to 4.2\% (19.7\% relative) and 5.1\% absolute (25.4\% relative) on the simulated and the replayed data over the pipelined system;
3) consistent performance improvements are also obtained using the proposed audio-visual multi-channel recognition system when even using occluded video input with the lip region randomly covered up to 60\%.

\noindent In summary, this work makes three main contributions:
\begin{itemize}
    \item This paper presents the first work on incorporating visual inputs in both the speech separation front-end and the recognition back-end within a bimodal and multi-channel inputs based overlapped speech recognition system. A systematic overview and comparison over three different audio-visual channel integration methods featuring a tight integration between the separation and the  recognition components is given. In contrast, video information is added into only either the separation front-end \cite{AVSE1, AVSEWJ, gu2019Multi}, or the recognition back-end alone \cite{AVSRJIAN, AVSR1,AVSR2016OVE}. A more holistic use of video cues as investigated in this paper was not considered. 
    \item This is the first work that uses an interpolated error cost that combines the lattice-free MMI based sequence discriminative training criterion and the scale-invariant signal to noise ratio (Si-SNR) metric to integrate the separation front-end and recognition back-end. In contrast, the previous research focused on using cross entropy based error cost \cite{MIMO2019,MIMO2020, JOINT} in the overall end-to-end system fine-tuning and integration stage. 
    \item This paper presents the first more complete attempt to investigate the effect from video input occlusion on both the separation and the  recognition components as well as the final system performance on overlapped speech. 
    In contrast, the previous investigation on the effect from occluded video is limited to either the separation module \cite{MYLIP} or recognition \cite{DFMSN2019AV} component.
\end{itemize}


The rest of the paper is organized as follows. Section 2 introduces three neural network based multi-channel integration methods. Section 3 presents various forms of audio-visual multi-channel speech separation networks. Description of the recognition back-end components and their integration with the separation front-end are given in section 4. Experimental results are presented in section 5. Section 6 draws the conclusions and discusses possible future directions.

\section{Multi-channel speech separation}

\subsection{Multi-channel signal model for overlapped speech}
\noindent Ignoring the reverberation in the overlapped speech, the spectrum of the received speech signal $X_r(t,f)$ recorded by a far-field microphone array composed of R channels can be modeled as:

\begin{align}
    X_r(t,f) = Y_r(t,f)+ N_r(t,f),
\end{align}
where $X_r(t,f)$, $Y_r(t,f)$ and $N_r(t,f)$ denote the short-time Fourier transform (STFT) spectra of the overlapped, target and interfering speech received by the $r$th microphone respectively. 
Without loss of generality, we select the first channel as the reference channel ($r=1$) in this paper.


\subsection{TF Masking}
\noindent To separate the target speaker from other interfering sources, the \textit{TF masking} approaches have been widely used in monaural speech separation tasks in the past few decades \cite{wang2008TF, WANG2018, CRM, TCN}.
Such approaches predict spectral TF mask labels that specify whether a particular TF spectrum point is dominated by the target speaker or other interfering sources to facilitate speech separation.
Recently, several researches have shown that integrating the multi-channel information collected by a microphone array can improve the mask estimation of the reference channel and lead to better speech separation. 
It has been found in previous research that the complex ratio masks (CRMs) outperform both the binary masks (BMs) and real-value ratio masks (RMs) on speech separation \cite{CRM2015,CRM} and enhancement \cite{hu2020dccrn} tasks. 
For this reason, the CRM based \textit{TF masking} approach is implemented in this work.
The complex spectrum of the separated output ${{Y}}(t,f)$ is computed as follows:
\begin{align}
    Y&(t,f)=M(t,f)X_{r}(t,f) \\
          &=\mathcal{R}\{M(t,f)\}\mathcal{R}\{X_{r}(t,f)\} - \mathcal{I}\{M(t,f)\}\mathcal{I}\{X_{r}(t,f)\} \nonumber\\  &+j(\mathcal{I}\{M(t,f)\}\mathcal{R}\{X_{r}(t,f)\} + \mathcal{R}\{M(t,f)\}\mathcal{I}\{X_{r}(t,f)\} \nonumber
\end{align}
where $M(t,f) \in \mathbb{C}$ is the CRM of the target speaker and $\mathcal{R}\{\cdot\}$/$\mathcal{I}\{\cdot\}$ denote the real/imaginary parts of a complex number respectively. 
Although the \textit{TF masking} approach can provide perceptually enhanced sounds, it has been reported that the artifacts resulting from deterministic spectral masking introduced a negative impact on downstream speech recognition system performance \cite{NTTchime3, ROM, MULTIFAROVER}.

\subsection{Multi-channel integration using beamforming}
\noindent The acoustic beamforming approaches are designed to capture sound coming from the target speaker direction while reducing interfering sounds coming from other directions. This is realized by setting the beamformer filter parameters to the target direction. 
A linear filter
\begin{align}
    {\bf{W}}(f)=[W_1(f), W_2(f), ..., W_R(f)]^{\text{T}} \nonumber
\end{align}
is applied to the multi-channel overlapped speech spectrum vector
\begin{align}
    {\bf{X}}(t,f)=[X_1(t,f), X_2(t,f), ..., X_R(t,f)]^{\text{T}} \nonumber
\end{align}
as follows:
\begin{align}
    Y(t,f) &= {\bf{W}}(f)^\text{H} {\bf{X}}(t,f) \nonumber \\
              &= \underbrace{{\bf{W}}(f)^\text{H} {\bf{Y}}(t,f)}_{\text{speech}} + \underbrace{{\bf{W}}(f)^\text{H} {\bf{N}}(t,f)}_{\text{noise}},
\end{align}
where $(\cdot)^{\text{H}}$ denotes the conjugate transpose.
The beamforming filter parameters in conventional beamformers are usually obtained by first estimating the steering vector, which requires the direction-of-arrival (DOA) of the target speaker before solving an optimization problem, such as MVDR beamformer.
With  the  successful  and  wider  application  of  deep  learning  based  speech  technologies, state-of-the-art neural beamforming techniques are represented by the following two approaches: 
1) using NNs to directly estimate beamforming filters as in \textit{Filter\&Sum} \cite{DEEPBEAMFORMER, FASNET1, FASNET2};  
2) using TF masks to estimate beamforming filters as in \textit{mask-based MVDR} or \textit{GEV} \cite{IMPMVDR, ROM, MULTIFAROVER}.

\subsection{Filter\&Sum}
\noindent The neural \textit{Filter\&Sum} beamforming approaches directly estimate the beamforming filter parameters in either  time domain \cite{Taramulti, FASNET1, FASNET2} or frequency domain \cite{DEEPBEAMFORMER} base on deep neural networks in a fully-trainable fashion. 
In this work, we adopt a frequency domain \textit{Filter\&Sum} approach to produce the separated output as follows:
\begin{equation}
    Y(t,f)={\bf{W}}(t,f)^\text{H} {\bf{X}}(t,f)=\sum_{r}W_{r}(t,f) * X_{r}(t,f).
\end{equation}
One limitation associated with the \textit{Filter\&Sum} beamformer is that the estimated filter parameters are allowed to change over very short analysis intervals, for example, between neighbouring frame windows of 25 milliseconds.  In practice this is an unrealistic assumption as the speech from a target speaker tends to remain from the same direction over a longer period  of time when collected using fixed microphone arrays, before he or she moves to a different position in the room.

\subsection{Mask-based MVDR}
\noindent When choosing the $r$th channel as the reference channel, the residual signal distortion $\xi_{r,d}(t,f)$ and the residual noise $\xi_n(t,f)$ can be computed by Equation (5) and Equation (6) respectively:
\begin{align}
    \xi_{r,d}(t,f) &= Y_r(t,f) - {\bf{W}}(f)^\text{H} {\bf{Y}}(t,f) \\
                     &= ({\bf{U}}_r-{\bf{W}}(f))^\text{H}{\bf{Y}}(t,f), \nonumber\\
    \xi_n(t,f) &= {\bf{W}}(f)^\text{H} {\bf{N}}(t,f)
\end{align}
where ${\bf{U_r}}=[0, 0, ..., 1 , ...,0]^T$ is a one-hot vector of which the $r$th entry equals to 1.
The MVDR beamformer is designed to minimize the noise output while imposing a distortionless constraint on the target speech signal \cite{MVDR}:
\begin{align}
    &\min_{{\bf{w}}(f)} E_{t}\{|\xi_n(t,f)|^2 \} \nonumber \\ \nonumber
    &\text{subject to}: \ E_{t}\{|\xi_{r,d}(t,f)|^2 \} = 0 
\end{align}
The distortionless constraint in the above optimization problem is equivalent to ${\bf{W}}(f)^\text{H}{\bf{G}}(f)=1$, which can be interpreted as maintaining the energy along the target direction. It can be shown that the solution of the above MVDR beamformer is:
\begin{align}
    {\bf{W}}(f)&=\frac{{\bf{\Phi}}_{n}{(f)}^{-1}{\bf{G}}(f)}{{\bf{G}}(f)^\text{H}{\bf{\Phi}}_{n}{(f)}^{-1}{\bf{G}}(f)} \\
    &=\frac{{\bf{\Phi}}_{n}{(f)}^{-1} {\bf{\Phi}}_{y}(f)}{{\text{Trace}}( {\bf{\Phi}}_{n}{(f)}^{-1} {\bf{\Phi}}_{y}(f)) } {\bf{U_r}},
\end{align}
where ${\bf{\Phi}}_{n}{(f)}= E_{t}\{ {\bf{N}}(t,f) {\bf{N}}(t,f)^\text{H}\}$ and ${\bf{\Phi}}_{y}{(f)}= E_{t}\{ {\bf{Y}}(t,f) {\bf{Y}}(t,f)^\text{H}\}$ are the PSD matrices of the noise and target speech respectively.
The MVDR filter parameter estimation in Equation (7) is expressed in terms of the noise PSD matrices and the steering vector. Alternatively it can also be re-expressed using both the target speech and noise PSD matrices as in Equation (8).

In \textit{mask-based MVDR} approaches, the deep neural networks are used to estimate the real-value\cite{ROM, MULTIFAROVER, MIMO2019} or complex\cite{CRM} TF masks of the target speech $M^{y}(t,f)$ and other interfering sources $M^n(t,f)$ respectively. The PSD matrices corresponding to each source can be calculated with the estimated TF masks shown as follows:
\begin{align}
    {\bf{\Phi}}_{y}(f)=\frac{\sum_{t=1}^{T} (M^{y}(t,f)*{\bf{X}}(t,f))({M^{y}(t,f)*\bf{X}}(t,f))^{\text{H}}}{\sum_{t=1}^{T} M^{y}(t,f)*(M^{y}(t,f))^{\text{H}}}, \nonumber\\
    {\bf{\Phi}}_{n}(f)=\frac{\sum_{t=1}^{T} (M^{n}(t,f)*{\bf{X}}(t,f))({M^{n}(t,f)*\bf{X}}(t,f))^{\text{H}}}{\sum_{t=1}^{T} M^{n}(t,f)*(M^{n}(t,f))^{\text{H}}}.
\end{align}
The MVDR beamformer filters can then be obtained using Equation (8). 
Compared with both the \textit{TF masking} and the \textit{Filter\&Sum} approaches, \textit{mask-based MVDR} beamformers using the spatial temporal correlation in the PSD matrices to enforce a consistent set of filter parameters to be estimated over the analysis window, in which the location of the speakers are unchanged. 
Hence, the processing artifacts observed in the former two approaches can be minimized. This is particularly useful when modelling the short speech segments within which the target speaker voice is recorded from the same direction using the array.
Compared with both the \textit{TF masking} and the \textit{Filter\&Sum} approach, 
the \textit{mask-based MVDR} approach retains a consistent DOA estimation with a beamforming analysis window over, for example, an utterance of speech and 
the minimum distortion constraint in traditional MVDR beamforming.
The \textit{mask-based MVDR} approach has demonstrated state-of-the-art performance in noisy and overlapped speech recognition \cite{ROM, MULTIFAROVER, MIMO2019}.

\begin{figure*}[htb]
\begin{subfigure}{1\textwidth}
  \centering
  \includegraphics[width=18cm]{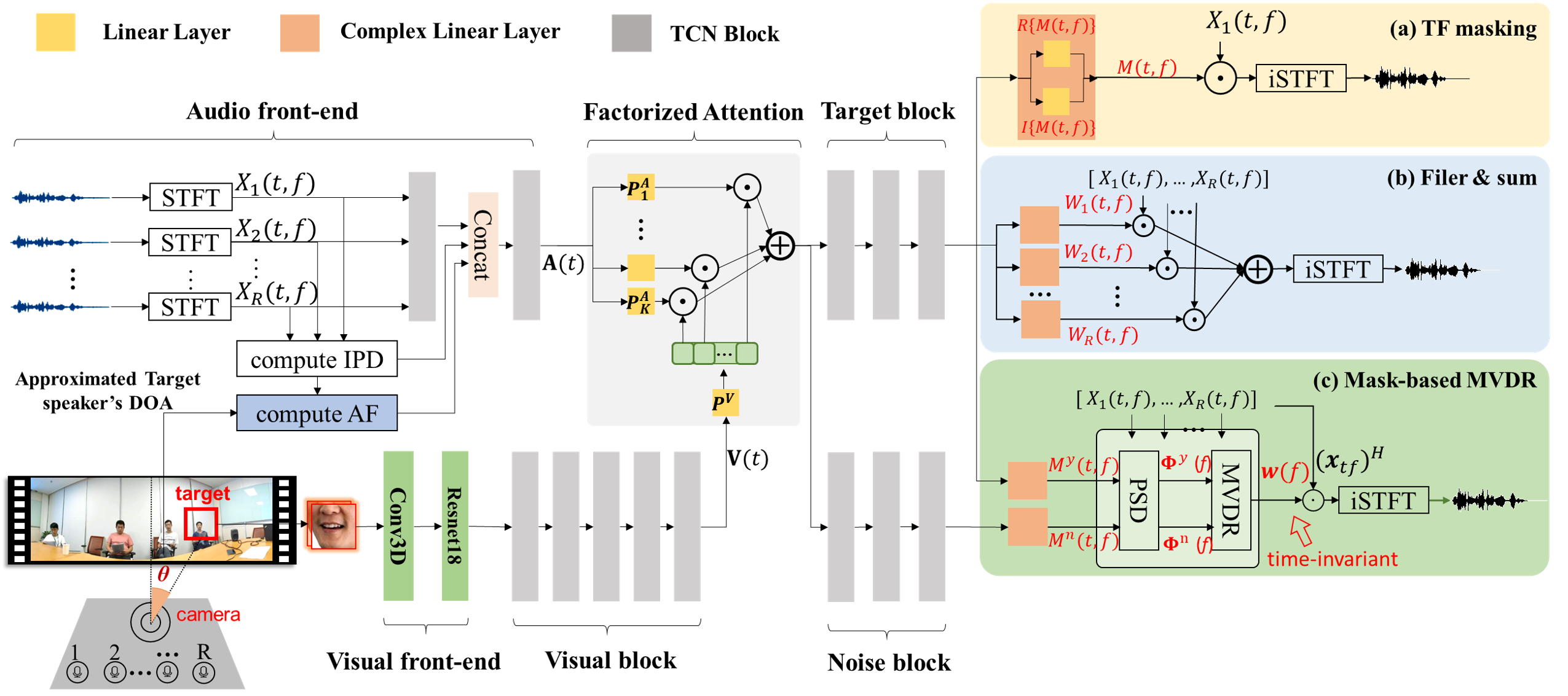}  
  \label{fig:sub-first}
\end{subfigure}
\caption{Illustration of the proposed audio-visual multi-channel speech separation networks, where  $X_{r}(t,f)$ is the complex spectrum of each channel. $\text{\bf V}(t)$ and $\text{\bf A}(t)$ denote the audio and the visual embedding at frame index $t$ respectively.
The detailed paradigm of the TCN block is demonstrated in Figure 2.
(a), (b) and (c) represent three options of channel integration approaches: (a) TF masking: $M(t,f)$ represents the complex mask of the target speaker, where $R\{(M(t,f)\}$) and $I\{(M(t,f)\}$ are the real and the imaginary part of the mask respectively; (b) Filter\&Sum: $W_{r}(t,f)$ denotes the beamforming filter parameters of the $r$th channel; (c) Mask-based MVDR: $M^{y}(t,f)$ and $M^{n}(t,f)$ are the complex masks of the target and the interfering sources, ${\bf{\Phi}}^{y}(f)$ and ${\bf{\Phi}}^{n}(f)$ are the corresponding PSD matrices and ${\bf{W}}(f)$ is the time-invariant beamforming filter parameters. }
\label{fig:fig}
\end{figure*}

\begin{figure}[htb]
  \centering
  \includegraphics[width=8cm]{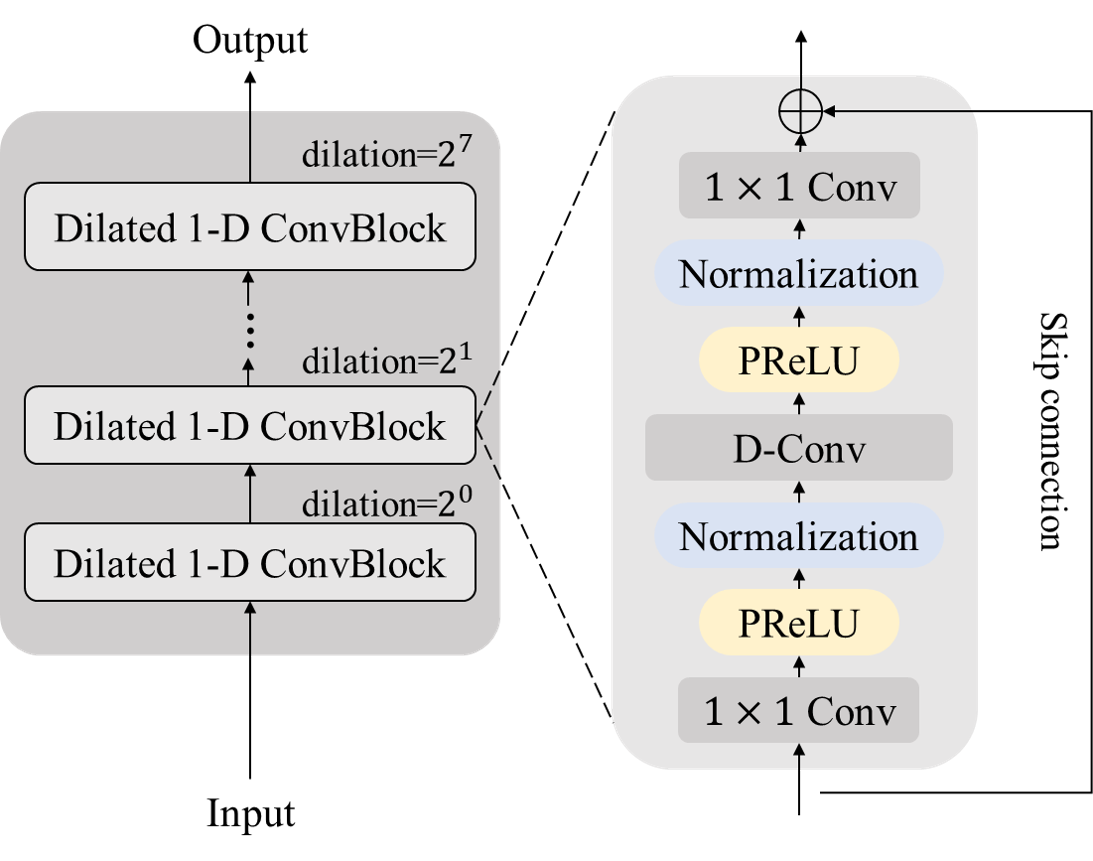}  
  \label{fig:sub-first}
\caption{Illustration of the architecture of the temporal convolutional network (TCN) Block. Each dilated 1-D ConvBlock consists of a 1$\times$1 convolutional layer, a depth-wise separable convolution layer (D--Conv) \cite{DCONV}, with PReLU \cite{PReLU} activation function and normalization added between each two convolution layers. Skip connection is added in each dilated 1-D ConvBlock.}
\label{fig:fig}
\end{figure}

\section{Audio-visual Multi-channel speech separation}
In this section, we introduce the proposed audio-visual multi-channel speech separation networks using \textit{TF masking}, \textit{Filter\&Sum} and \textit{mask-based MVDR} channel integration methods. 

\subsection{Audio modality}
In the proposed separation front-ends, three types of audio features including the complex spectrum, the inter-microphone phase differences (IPDs) \cite{ROM} and the location-guided angle feature (AF) \cite{Angle, gu2019Multi} are adopted as the audio inputs.  
The detailed paradigm of the audio modality processing is illustrated in the top left corner of Figure 1.
The complex spectrum of all the microphone array channels are first computed through the STFT.
%
Following the same recipe as in \cite{gu2020multi}, the IPD feature is calculated as follows: 
\begin{equation}
    {\rm{IPD}}^{(i,j)}(t,f)=\angle\Bigg{(}\frac{X_{i}(t,f)}{X_{j}(t,f)}\Bigg{)},
\end{equation}
where $X_{i}(t, f)$ represents the i-th channel's complex spectrum of the mixed signal at time frame $t$ and frequency bin $f$, $(i, j)$ corresponding to the selected microphone pair and $\angle (\cdot)$ outputs the angle of the input argument. 
The IPD feature captures the relative phase difference between microphones, which is correlated with the time difference of arrival (TDOA).

In addition, when the geometry of the microphone array and the direction of arrival (DOA) of the target speaker $\theta$ are given, the steering vector corresponding to the target speaker can be computed as:

\begin{align}
    {\bf{G}}(f) =  [e^{-j\frac{2\pi f d_{11}}{c}\cos(\theta)}, e^{-j\frac{2\pi f d_{1r}}{c}\cos(\theta)}, ...,e^{-j\frac{2\pi f d_{1R}}{c}\cos(\theta)}]
\end{align}
where $d_{1r}$ is the distance between the  first (reference) and $r$th microphone ($d_{11} = 0$). $c$ is the sound velocity.

Based on the computed steering vector,  the location-guided AF feature introduced in \cite{Angle,gu2020multi} are also adopted to provide discriminative information for the target speaker as follows:
\begin{align}
    \text{AF}(t,f) = \sum_{\{(i,j)\}} \frac{\big{\langle}\text{vec}\big{(}\frac{G_j(f)}{G_i(f)}\big{)}, \text{vec}\big{(}\frac{X_{i}(t,f)}{X_j(t,f)}\big{)}\big{\rangle}}{\big{\|}\text{vec}\big{(}\frac{G_j(f)}{G_i(f)}\big{)}\big{\|}\cdot \big{\|}\text{vec}\big{(}\frac{X_{i}(t,f)}{X_j(t,f)}\big{)}\big{\|}}
\end{align}
where $\|\cdot\|$ denotes the vector norm, $\langle\cdot,\cdot\rangle$ represents the inner product and $\{(i,j)\}$ denotes the selected microphone pairs.
$\text{vec}(\cdot)$ transforms the complex value into a 2-D vector, where the real and imaginary parts are regarded as the two vector components. 
The design principle of the AF is that if the TF bin $X_i(t,f)$ is dominated by the target speaker from direction $\theta$, then its corresponding $\text{AF}(t,f)$ will be close to 1, otherwise close to 0.
In this work, the DOA of the target speaker can be estimated by tracking the speaker's face from a 180-degree wide-angle camera as shown in Figure 1 (bottom left corner).

Motivated by the success of Conv-Tasnet \cite{TCN} in speech separation, the temporal convolutional network (TCN) architecture, which uses a long reception field to capture more sufficient contextual information, is adopted in our separation front-ends. 
As shown in Figure 2, each TCN block is stacked by 8 Dilated 1-D ConvBlock with exponentially increased dilation factors $2^0, 2^1, .... ,2^7$.
As shown in the Audio front-end (Figure1, top left corner), the complex spectrum of each microphone array channel are first concatenated and then fed into a TCN block. 
The outputs are concatenated with the IPD and AF features and then fed into another TCN block to compute the audio embeddings ${\bf{A}}\in\mathbb{R}^{T\times D}$.

\subsection{Visual modality}
For the visual modality, as shown in the bottom left corner of Figure 1, the lip region of the target speaker obtained by face tracking is fed into the Visual front-end (Figure 1, bottom left corner in green) followed by the Visual block (Figure 1, bottom middle in gray) to compute the visual embeddings ${\bf{V}}\in\mathbb{R}^{T\times D}$. 
The network structure of the Visual front-end is similar to the one proposed in \cite{LIPNET}, which consists of a spatio-temporal convolution layer (Conv3D) and a 18-layer ResNet \cite{ResNet} to capture the spatio-temporal dynamics of the lip movements. 
The Visual block consists of 5 TCN blocks.
Following the work in \cite{AVSRJIAN,YU2020AV, AVSRHYBRIDCTC}, the Visual front-end is pretrained on the lipreading task as described in \cite{LIPNET}.
The visual modality can provide not only discriminative information to facilitate phone classification, but also crucially additional cues to track and separate the target speaker from interfering sources of sound.

\subsection{Modality fusion}
In order to effectively integrate the audio and visual modalities, a careful design of the modality fusion scheme is required. 
Based on the investigation of different modality fusion methods in our previous work \cite{gu2019Multi}, a factorized attention-based modality fusion method, which has been proven to outperform the most commonly used feature concatenation method \cite{AVSE1, L2L2018, AVSEWJ, MYLIP} in the audio-visual speech separation front-ends, is adopted in this work.

As shown in Figure 1 (middle up, in light gray), the acoustic embedding  ${\bf A}(t)$ at frame index $t$ is first factorized into $K$ acoustic subspace vectors by a series of parallel linear transformations ${\bf{P}}^{A}_{k}\in\mathbb{R}^{D\times D}$ and the visual embedding ${\bf V}(t)$ is mapped into a $K$ dimensional vector ${\bf{v}}(t) = [v_{1}(t), ..., v_{K}(t)]$ by projection matrix ${\bf{{P}}^{V}}\in\mathbb{R}^{D\times K}$ in the factorized attention method as follows: 
\begin{align}
    &[{\bf{a}}_{1}(t), ..., {\bf{a}}_{K}(t)] = [{\bf{P}}^{A}_{1}, ..., {\bf{P}}^{A}_{K}] {\bf{A}}(t) \\
    &{\bf{v}}(t) = \text{Softmax}\Big{(}{\bf{{P}}^{V}}{\bf{V}}(t)\Big{)}.
\end{align}
Then the fused audio-visual embedding ${\bf AV}(t)\in\mathbb{R}^{D}$ is obtained by using the weighted sum of the acoustic subspace vectors: 
\begin{align}
    {\bf{AV}}(t) = \sigma\Big{(}\sum_{k=1}^{K}v_{k}(t) {\bf{a}}_{k}(t) \Big{)}
\end{align}
where $\sigma(\cdot)$ is the sigmoid function.


\subsection{Channel integration}
As discussed in section II, three different audio-visual multi-channel integration approaches are investigated in this work.

\noindent a) {\bf \textit{TF masking}}: The diagram of the \textit{TF masking} approach is illustrated in Figure 1 (top right, in light yellow).
The hidden outputs of the Target block (Figure 1, middle up in gray) are fed into a complex linear layer to estimate the complex mask of the reference channel.
The structure of the complex linear layer is shown in Figure 1 (top right in orange), which consists of two linear layers. One is used to estimate the real part $\mathcal{R}\{M(t,f)\}$ of the complex mask, the other is used to estimate the imaginary part $\mathcal{I}\{M(t,f)\}$.
Based on the estimated TF mask, the output complex spectrum is then computed via Equation 2.


\noindent b) {\bf \textit{Filter\&Sum}}: The diagram of the \textit{Filter\&Sum} approach is shown in Figure 1 (right middle, in light blue). 
Different from the \textit{TF masking} approach, the hidden outputs of the Target block are fed into a series of complex linear layers to estimate the time variant beamforming filter parameters $W_{r}(t,f)$ corresponding to each channel frame by frame. 
The frequency domain beamforming outputs are then computed using Equation 4.

\noindent c) {\bf \textit{Mask-based MVDR}}: The \textit{mask-based MVDR} approach is demonstrated Figure 1 (right bottom, in green). 
Different from the \textit{TF masking} and the \textit{Filter\&Sum} approaches, an additional Noise block (Figure 1, middle bottom in gray) containing 3 TCN blocks and a complex linear layer is adopted to estimate the complex TF mask $M^{n}(t,f)$ for the noise signals.
As discussed in \cite{ROM}, estimating the TF masks for both the target and noise signals can improve the speech separation performance of the \textit{mask-based MVDR} approach.
With the TF masks of the target and interference speech, the beamforming filter parameters are calculated using Equation (8) and (9)  described in section II-E.
In this work, we assume that the location of the speakers are unchanged within each utterance, which is common in meeting and restaurant environment. 
Therefore, in the \textit{mask-based MVDR} approach, the beamforming filter parameters ${\bf{W}}(f)$ are fixed with a beamforming analysis window, for example, an utterance of speech in this work.

In all the three channel integration methods, the target speech complex spectrum extracted by each channel integration method is used to compute the target speech waveform using the inverse STFT (iSTFT) operation.


\section{Audio-visual multi-channel speech recognition}
In this section, we first introduce our audio-visual speech recognition back-ends and then describe the approaches to integrate the separation and the  recognition components.

\subsection{Audio-visual speech recognition back-end}
Extensive audio-visual speech recognition technologies have been conducted in recent years and demonstrated their efficacy in improving speech recognition performance under both clean and adverse conditions \cite{AVSRBEGIN, AVSR2004, AVSR2013, AVSR2015, AVSR2020HAN, AVSR2016OVE, AVSR1, DFMSN2019AV}.
Following \cite{YU2020AV}, in this work, the convolutional long short-term memory fully connected neural network (CLDNN) \cite{CLDNN} is adopted as the recognition back-end system architecture.
As shown in Figure 3 (left, in dark gray), the log filter bank features are first calculated from the separated target speech waveform before being concatenated with the visual features extracted using the Visual front-end. 
The concatenated features are fed into the CLDNN network to estimate the frame level posteriors.
To optimize the model parameters in the recognition back-end, two widely used training criteria i.e. CTC \cite{CTC} and LF-MMI \cite{LFMMI, FLATLFMMI} are investigated in this work:

\noindent 1) {\textbf{CTC}}: The CTC approach uses a blank symbol, which can appear between the modelling units (graphemes, phonemes), to define an objective function that sums over all possible alignments of the reference transcription with the input sequence of speech frames:
\begin{align}
    \mathcal{L}_{CTC} &= \sum_{u=1}^{U} \log \sum_{{\pmb{\pi}}^{u}, {{\pi}}_{t}^{u}\in\{{\bf \Omega}\cup\epsilon\}} \prod_{t=1}^{T} P (\pi_{t}^{u} |{\bf O}^{u}) 
\end{align}
where ${\bf O}^{u}=[O^{u}_1, ..., O^{u}_T]$ represents the input utterance of $T$ frames and
$\bf \Omega$ denotes the grapheme (phoneme) symbol set. 
${\pmb \pi}^{u}=[\pi^{u}_{1}, ..., \pi^{u}_{T}]$ represents
a possible alignment between ${\bf O}^{u}$ against the CTC output token $\pi^{u}_{t}$, which are based on either a grapheme (phoneme) symbol, or a special null emission “$\epsilon$“ token, as considered in this paper.


\noindent 2) {\textbf{LF-MMI}}:
Sequence discriminative training techniques, represented by lattice-free MMI \cite{LFMMI}, have defined state-of-the-art hybrid ASR system performance in the past few years.
The MMI criterion is a discriminative objective function which aims to maximize the probability of the reference transcription while minimizing the probability of all other transcriptions:
\begin{align}
                     \mathcal{L}_{MMI} = \sum_{u=1}^{U} \log \frac{P({\bf{O}}|{\bf{H}}^{u})P({\bf{H}}^{u})}{\sum_{\tilde{\bf{H}}^{u}}P({\bf{O}}|{\bf{\tilde{H}}}^{u})P({\bf{\tilde{H}}}^{u})} \nonumber
\end{align}
\noindent where ${\bf{\tilde{H}}}^{u}$ represents any possible transcriptions. 
In recent research \cite{FLATLFMMI, E2ELFMMI}, it has been shown that the end-to-end LF-MMI approach 
can outperform CTC based approach using either phoneme or grapheme modelling units on clean speech.

\subsection{Integration of the separation \& recognition components}

\begin{figure}[ht]
\begin{minipage}[b]{1\linewidth}
  \centering
  \centerline{\includegraphics[width=9cm]{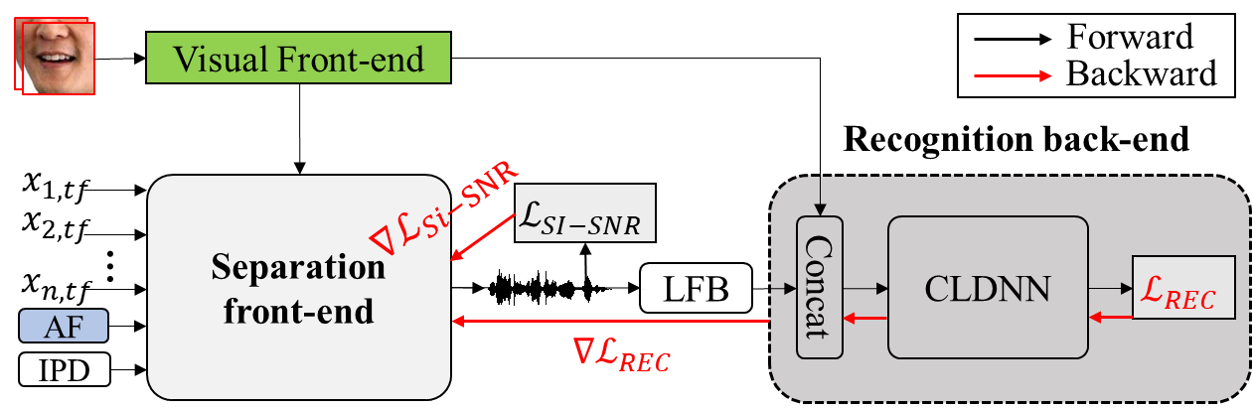}}
\end{minipage}
\caption{Joint fine-tuning: $\nabla\mathcal{L}_{REC}$ and $\nabla\mathcal{L}_{Si-SNR}$ represent the gradients of speech recognition i.e CTC, LF-MMI and speech separation Si-SNR objective functions respectively, "LFB" denotes log filter bank acoustic features.}
\label{figures}
\end{figure} 
Traditionally, the speech separation and recognition components are developed separately and then used in a pipelined fashion \cite{IMPMVDR, ROM, Angle, MULTIFAROVER}. 
However, two issues arise with such approach: 
1) the cost function mismatch between separation and recognition components cannot guarantee the separated outputs target to optimal recognition performance; 
2) the artifacts created by separation can increase modeling confusion of the recognition component and lead to performance degradation. 

According to \cite{Beamnet,CRM, JOINT, MAX2019}, tight integration of the two components with joint ﬁne-tuning can address above two issues. 
In this work, we investigated three variants of fine-tuning methods: 
1) fine-tuning the recognition system only on the enhanced signals;
2) jointly fine-tuning the separation and the  recognition components using the recognition cost function; 
3) jointly fine-tuning both systems using a multi-task criterion, which interpolates the recognition and Si-SNR cost functions:
\begin{equation}
    \mathcal{L} = \mathcal{L}_{REC} + \alpha \mathcal{L}_{Si-SNR},
\end{equation}
where $\alpha$ is a manually tuned weight of the Si-SNR loss and $\mathcal{L}_{REC}$ can be either $\mathcal{L}_{CTC}$ or $\mathcal{L}_{LF-MMI}$ cost function. As shown in Figure 2, the gradient of the recognition cost is propagated into the separation front-end to update the model parameters of the entire system.

\section{EXPERIMENT SETUP}
In this section, we first introduce the details of the corpus adopted in this work. 
Second, we describe the details of generation process of the multi-channel overlapped speech by either simulation or replay.
Third, we explain how we introduce visual occlusion into the video.
Finally, we introduce the implementation details of the proposed systems.

\subsection{LRS2 corpus}
The Oxford-BBC Lip Reading Sentences 2 (LRS2) corpus \cite{LRS2}, which is one of the largest publicly available corpora for audio-visual speech recognition, is adopted in our experiments.
This corpus consists of news and talk shows from BBC program, which is a challenging task since it contains thousands of speakers with large variation in head pose.
The LRS2 corpus is divided into three subsets, i.e. {\tt Pre-train}, {\tt Train-val} and {\tt Test} set.
In our experiments, the Pre-train and Train-Val subsets are combined for system training.
More statistic details of the LRS2 corpus can be found in \cite{AVSR1}.

\subsection{Simulated overlapped speech}
Since there is no publicly available audio-visual multi-channel overlapped speech corpus, we simulated the multi-channel overlapped speech in our experiments based on the LRS2 corpus. 
Details of the simulation process is described in Algorithm 1.
A 15-channel symmetric linear array with non-even inter-channel spacing is used in the simulation process, as shown in Figure 4.
Reverberation is also added in the simulated data by convolving the single channel signals with the Room Impulse Responses (RIRs) generated by the image-source method \cite{IMAGE}.
The room size is randomly selected ranging from 4$\times$4$\times$2.5 m$^3$ to 10$\times$8$\times$6 m$^3$ (length$\times$width$\times$height) and the reveberation time T60 is sampled from a range of 0.05 to 0.7s.
The average overlapping ratio of the simulated utterances is around 85\% and SIR is around 0dB.
The simulated data is divided into three subsets with 14.2k, 4.6k and 1.2k utterances respectively for training (200h), validation (2h) and evaluation (0.5h).
\begin{figure}[htb]
    \centering
    \includegraphics[width=9cm]{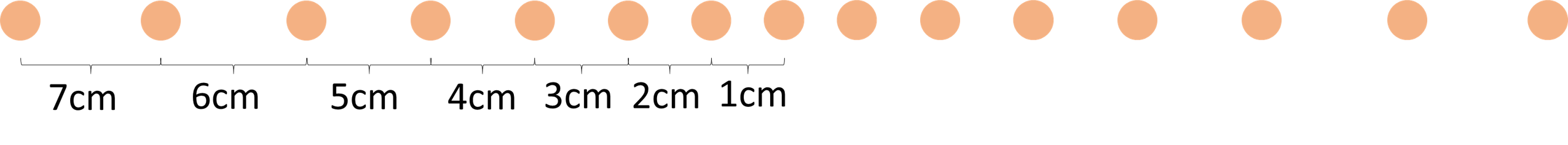}
    \caption{The architecture of the microphone array used in the simulation and replay data recording.}
    \label{fig:my_label}
\end{figure}
\begin{algorithm}
    \caption{Data simulation process of multi-channel overlapped speech }
    {\bf{Input}:} single channel non-overlapped LRS2 speech
    \begin{algorithmic}[1]
        \For{utterance in LRS2}
        \State {Randomly select an interfering utterance from another speaker in LRS2 corpus}
        \State Sample a SIR uniformly from (-6,0,6) dB
        \State Randomly generate microphone array and speakers' position (Distance between speakers and array is 1-5m) 
        \State Scale the target and interferring sources with the sampled SIR
        \State Generate mixed speech per channel with overlapping ratio randomly from 60\% to 100\%
        \EndFor
    \end{algorithmic}
    {\bf{Output}:} multi-channel overlapped speech
\end{algorithm}

\begin{figure}[h]
\begin{minipage}{.5\textwidth}
\centering
\includegraphics[width=1\textwidth]{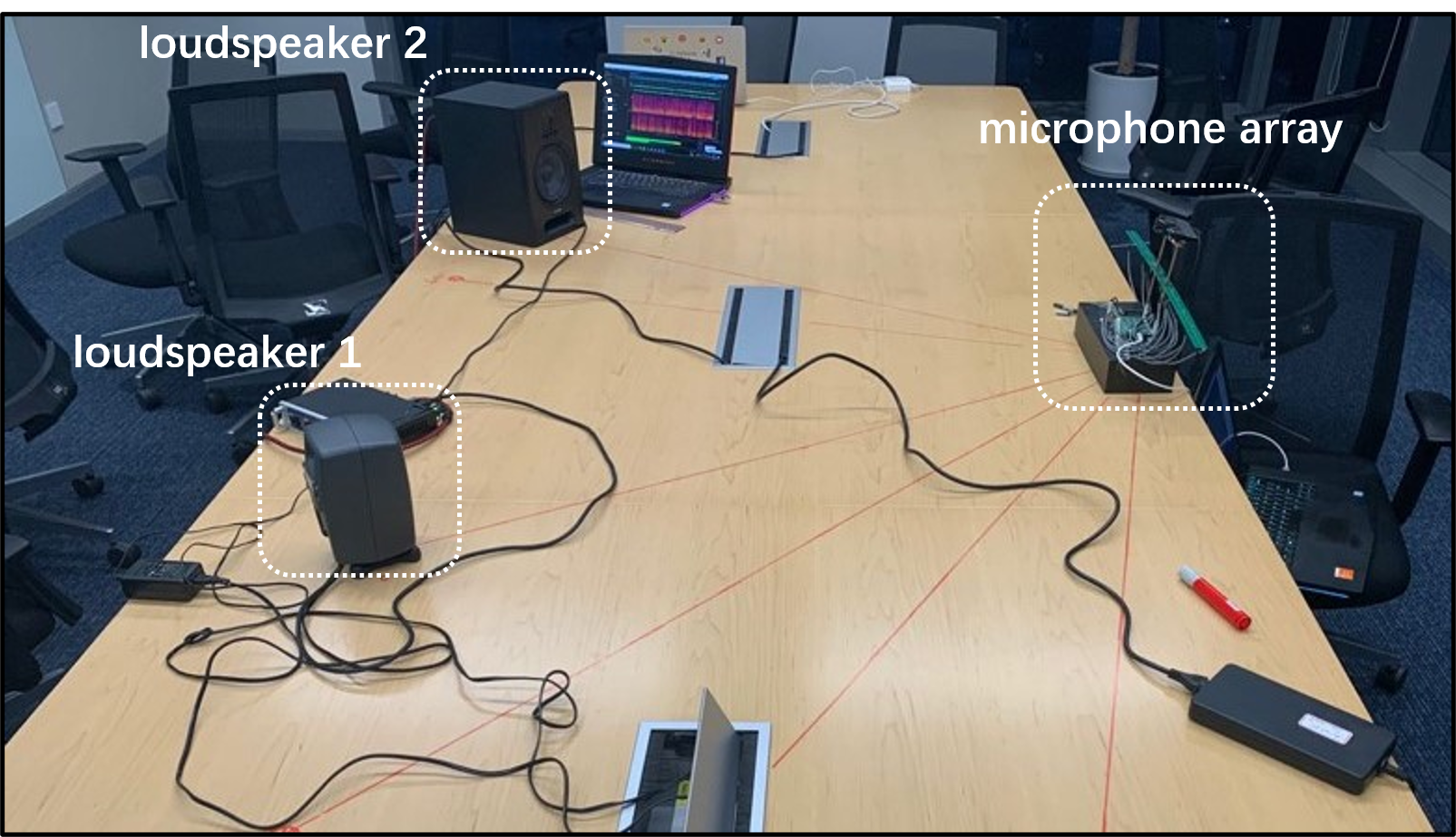}
\label{fig:fig1}            
\end{minipage}%

\caption{Replayed recording of overlapped LRS2 test set}
\end{figure}
\subsection{Replayed overlapped speech}
To further evaluate the performance of the proposed systems in realistic environment, a replayed test set with 1.2k (0.5h) utterances recorded in a 10$\times$5$\times$3 m$^3$ meeting room is also used in our experiments.
As shown in Figure 5, two loudspeakers are used to replay different utterances of the LRS2 test set simultaneously to generate overlapped speech.
The structure of the microphone array used during recording is the same as that used in simulation.
The target and interfering speakers are located at following directions related to the mounted camera, i.e. 
(15$^\circ$,30$^\circ$), (45$^\circ$,30$^\circ$), (75$^\circ$,30$^\circ$), (105$^\circ$,30$^\circ$), (30$^\circ$,60$^\circ$), (90$^\circ$,60$^\circ$), (120$^\circ$,60$^\circ$) and (150$^\circ$,60$^\circ$), 
where the distance between the loudspeakers and microphones ranges from 1m to 1.5m.
In the replayed data, the approximated DOA of the target speaker is obtained by the 180 degree camera mounted on top of the microphone array.
The average overlapping ratio of the replayed overlapped speech is around 80\% and SIR is around 1.5dB.

\subsection{Visual occlusion and low-resolution}
As discussed in Section I, the performance of audio-visual speech separation and recognition systems crucially depends on the quality of the video input fed into these systems. 
In the experiments of this paper, an ablation study is conducted to assess the impact on system performance due to two forms of video input quality degradation often found in real world applications, before a set of techniques designed to improve robustness against such degradation are evaluated and their performance analysed. 
First, low-resolution visual inputs were generated by gradually reducing the original video resolution from $160\times160$ pixels to $120\times120$, $80\times80$, $60\times60$, $40\times40$, $30\times30$, $20\times20$, and eventually down to $10\times10$ pixels , as shown in Figure 6 (a).
Second, the video frames were occluded by applying randomly sized and positioned square patches to the lip region of each speaker. The size of the occluded lip regions randomly varies from $45\times45$ to $60\times60$ pixels. 
For each utterance, occlusion is applied to a randomly sampled window of consecutive video frames. 
The ratio between uncovered and occluded frames is randomly sampled from the following settings: 20\%, 40\%, 60\% and 80\%. As such video occlusion is applied to a contiguous area of the lip region, it is regarded as a more naturalistic form of video quality degradation in comparison with the frame level drop-out approach \cite{DFMSN2019AV,gu2019Multi}.


\begin{figure}[htb]
     \centering
     \begin{subfigure}[b]{0.49\textwidth}
         \centering
         \includegraphics[width=7cm]{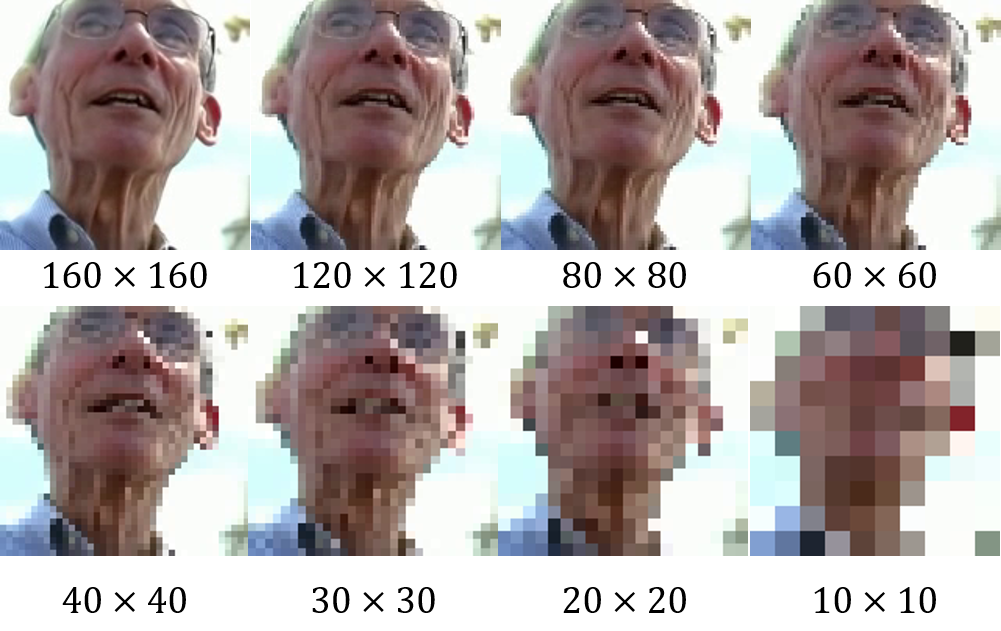}
         \caption{Low-resolution visual inputs}
         \label{fig:y equals x}
     \end{subfigure}
     \hfill
     \begin{subfigure}[b]{0.49\textwidth}
         \centering
         \includegraphics[width=8.8cm]{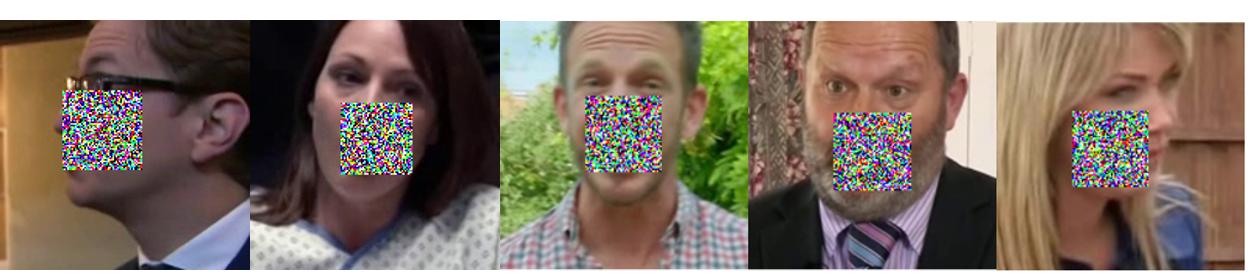}
         \caption{Visual occlusion}
         \label{fig:three sin x}
     \end{subfigure}
        \caption{Examples of (a) low-resolution visual input by gradually reducing the original video resolution from $160\times160$ pixels  to $10\times10$ pixels; (b) occluded visual input with randomly sized ($45\times45$ to $60\times60$ pixels) and positioned square patches applied to the lip region.  }
        \label{fig:three graphs}
\end{figure}

\subsection{Implementation details}
\noindent{\bf{Features:}}
1) In the separation front-ends (Figure 1, top left corner), the 257-dimensional complex spectrum of each channel is extracted using a 512-point FFT with 32ms hanning window and 16ms frame rate.
In our implementation, the STFT operation is implemented as a convolution layer to enable on-the-fly computation.
The AF and IPD features are computed using 9 microphone pairs (1,15), (2, 14), (3, 13), (1, 7), (12, 4), (11, 5), (12, 8), (7, 10) and (8, 9). 
These microphone pairs are selected to sample different spacing between microphones following \cite{MULTIBAND, gu2020multi}.
2) The 40-dimensional log filter bank features extracted using a 40ms window and 10ms frame rate are adopted as the input feature of the recognition back-end. Similar to the STFT operation, the log filter bank extractor is also implemented as a layer in the network to enable on-line extraction. 
3) For the visual front-end, the original 160$\times$160 video frames in LRS2 are centrally cropped by a 112$\times$112 window and then up-sampled to align with the audio frames via linear interpolation.

\noindent{\bf Separation front-end:}
In the separation front-ends (Figure 1, middle, in gray), for each TCN block (Figure 2), the number of channels in the $1\times1$ Conv layer is set to be 256 for every Dilated 1-D ConvBlock. 
As for the D-Conv layer, the kernel size is set to be 3 with 512 channels.
The Visual front-end (Figure 1, bottom left corner in green) uses the same hyper-parameter settings as described in \cite{LIPNET}.
Following \cite{gu2020multi}, the number of the acoustic subspace $K$ is set to be 10 with ${\bf {\bf{P}}^{V}} \in \mathbb{R}^{256\times10}$ and  ${\bf P}_{k}^{A} \in \mathbb{R}^{256\times256}$ in the factorized attention layer.
The output dimension of the complex linear layer is set to be 257.

\noindent{\bf Recognition back-end}:
In our experiments, the CTC and LF-MMI based recognition back-ends use the same neural network structure, which consists of four 2-dimensional convolutional layers with channel sizes (64, 64, 128, 128) and  kernel size $3{\rm{x}}3$ followed by four 1280 hidden units BLSTM layers and a softmax layer.
Context-free grapheme units are used as the output layer targets in both the CTC and LF-MMI based models.
The end-to-end LF-MMI criterion is implemented following the recipe\footnote{https://github.com/pytorch/examples/} in \cite{pychain}.
The language model (LM) used in recognition is a 4-gram LM constructed on 2.33M words of the LRS2 {\tt Train-val} and {\tt Pre-train} data transcripts.

All of our models are trained using 4 NVIDIA Tesla P40 GPU cards. 
For all results presented in this paper, matched pairs sentence-segment word error (MAPSSWE) based statistical significance test was performed at a significance level $\alpha$= 0.05.

\section{Experimental Results}
In this section, we describe the experiment results.
First, to investigate the effectiveness of visual features extracted from the video frames, we compare the audio-only and audio-visual speech recognition systems without explicit speech separation components on non-overlapped and overlapped speech.
Second, to tightly integrate the separation front-end and recognition back-end, we investigate the performance of three different integration methods in the proposed systems. 
We use the original LRS2 utterances as the echo free non-overlapped speech.
The reverberant non-overlapped speech is simulated from the original LRS2 data using image-source method.
Third, we systematically investigate the impact of the visual features on the proposed system to confirm the strength and importance of the visual information.  
Finally, we investigate the impact of visual occlusion on the proposed systems.

\subsection{Speech recognition without separation front-end}
Table I presents the WER results of the  CTC and LF-MMI based ASR and AVSR systems without using microphone array and explicit speech separation components on echo free and reverberant speech with or without speech overlapping.

\noindent Several trends can be observed from Table I:
\begin{enumerate}
    \item For both the CTC and LF-MMI based systems, using visual information can significantly improve the recognition performance over the audio-only systems by up to 1.27\% (sys.1 vs. sys.2) and 2.75\% (sys.7 vs. sys.8) absolute WER reduction on echo free and reverberant non-overlapped speech.
    Especially, the audio-visual recognition system largely outperforms the audio-only system by up to 33.28\% and 48.62\% (sys.9 vs. sys.10) absolute WER reduction on simulated and replayed overlapped speech respectively,  which proves the effectiveness of the extracted visual features on overlapped speech recognition.
    \item In our experiments, both the reverberation and the interfering speech are introduced into the simulated and the replayed multi-channel overlapped speech. Compared with the large performance degradation over 50\% absolute WER increase caused by speech overlapping (sys.5 vs. sys.9, sys. 7 vs. sys.11), the reverberation only introduces around 4\% absolute WER degradation against the echo free speech (sys.1 vs. sys.5, sys.3 vs. sys.7). 
    This indicates that overlapping speech (sys.9-12) is the more dominant contributing factor leading to large performance degradation against clean speech based recognition systems (sys.1-4)  than reverberation (sys.5-8) on the LSR2 data considered in this paper.
    \item The LF-MMI based systems outperform the CTC based systems on both non-overlapped and overlapped speech with and without visual modality in our experiments. 
\end{enumerate}
Based on the second observation, we focus on solving the speech overlapping issue in this work. 
Since we are not aiming at dereverberation in our overlapped speech recognition systems, the WER results on the reverberant non-overlapped speech (sys.5-8) can be defined as the upper bound for all subsequent experiments. 

\begin{table}[htb]
    \caption{Performance of single channel ASR and AVSR systems on echo free and reverberant speech with or without overlapping. "simu" and "replay" denotes the simulated and the replayed test data. $\dagger$ denotes a statistically significant improvement is obtained over the corresponding ASR baseline.  }
    \label{tab:my_label}
    \centering
    {\scalebox{1}{
    \begin{tabular}{c|c|c|c|c|c}
    \toprule
    \multirow{2}{*}{Sys} &\multirow{2}{*}{Data} & \multirow{2}{*}{Criterion} & \multirow{2}{*}{+visual}   & \multicolumn{2}{c}{WER (\%)} \\
    \cline{5-6}
    &&&&simu&replay \\
    \hline
    1& & \multirow{2}{*}{$\mathcal{L}_{CTC}$}& \xmark   &\multicolumn{2}{c}{11.04} \\
    2&Echo free && \cmark      &\multicolumn{2}{c}{9.77$^\dagger$}          \\
    \cline{3-6}
    3& non-overlapped&\multirow{2}{*}{$\mathcal{L}_{LF-MMI}$} &\xmark   &\multicolumn{2}{c}{{9.44}} \\
    4& && \cmark       &\multicolumn{2}{c}{{8.55$^\dagger$}}             \\
    \hline
    \hline
    5& &\multirow{2}{*}{$\mathcal{L}_{CTC}$}& \xmark &\multicolumn{2}{c}{15.33} \\
    6&Reverberant& &\cmark    &\multicolumn{2}{c}{{13.93$^\dagger$}} \\
    \cline{3-6}
    7&non-overlapped&\multirow{2}{*}{$\mathcal{L}_{LF-MMI}$}& \xmark  &\multicolumn{2}{c}{{14.36}} \\
    8&&& \cmark      &\multicolumn{2}{c}{{11.61$^\dagger$}} \\
    \hline
    \hline
    9&&\multirow{2}{*}{$\mathcal{L}_{CTC}$}& \xmark  & 75.34 & 80.55 \\
    10&raw channel 1& &\cmark    &32.06$^\dagger$ & 31.93$^\dagger$ \\
    \cline{3-6}
    11&overlapped&\multirow{2}{*}{$\mathcal{L}_{LF-MMI}$}& \xmark  &65.44 &71.03 \\
    12&& &\cmark       &28.92$^\dagger$ & 28.89$^\dagger$ \\
    \bottomrule
    \end{tabular}}
    }
\end{table}


\subsection{Performance of audio-visual speech separation front-ends}
The Si-SNR results of the \textit{TF-masking}, \textit{filter\&sum} and \textit{mask-based MVDR} separation front-ends are shown in Table II.
Several trends can be observed in Table II:
\begin{enumerate}
    \item Using visual features in the separation front-ends can improve the the Si-SNR performance (sys.1 vs. sys.3).
    \item Separation front-ends using only visual features have comparable results with separation front-ends using only AFs.
    \item Compared with the \textit{TF-masking} and \textit{filter\&sum} separation front-ends, the \textit{mask-based MVDR} separation front-ends show relatively lower Si-SNR. One possible explanation is that the \textit{mask-based MVDR} benefits from the distortionless constraint which is not adopted in the other two approaches. 
\end{enumerate}
Figure 7 shows example spectra of target clean, overlapped, audio-only separated, and audio-visual separated speech segments obtained using the \textit{TF masking} based speech separation front-end. The spectrum portions circled using yellow dotted lines in (c) and (d) represent the  interfering speaker’s speech, which is largely removed in (d) after applying audio-visual speech separation\footnote{More examples of audio-visual multi-channel speech separation can be found in: https://yjw123456.github.io/Audio-visual-Multi-channel-Integration-and-Recognition-of-Overlapped-Speech/}.

\begin{table}[htb]
    \centering
    \caption{Si-SNR results of \textit{TF-masking}, \textit{filter\&sum} and \textit{mask-based MVDR} separation front-ends.}
    \begin{tabular}{c|cc|cccc}
    \toprule
        Sys & AF        & +visual          & TF masing & Filter\&Sum & MVDR \\
        \hline
        1&  \cmark   & \xmark           & 9.40     & 10.87      & 8.73\\
        2&  \xmark   & \cmark           & 9.77     & 11.02      & 8.84\\
        3&  \cmark   & \cmark           & 10.16    & 11.60      & 9.03\\
    \bottomrule
    \end{tabular}
    \label{tab:my_label}
\end{table}

\begin{figure}[htp]
     \centering
     \begin{subfigure}[b]{0.49\columnwidth}
         \centering
         \includegraphics[width=1\textwidth]{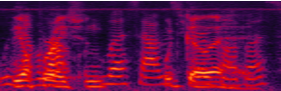}
         \caption{Overlapped}
         \label{fig:y equals x}
     \end{subfigure}
     \hfill
     \begin{subfigure}[b]{0.49\columnwidth}
         \centering
         \includegraphics[width=1\textwidth]{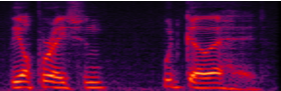}
         \caption{Target clean}
         \label{fig:three sin x}
     \end{subfigure}
     \hfill
     \begin{subfigure}[b]{0.49\columnwidth}
         \centering
         \includegraphics[width=1\textwidth]{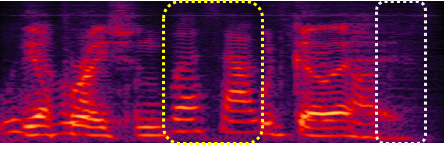}
         \caption{Audio-only}
         \label{fig:five over x}
     \end{subfigure}
     \begin{subfigure}[b]{0.49\columnwidth}
         \centering
         \includegraphics[width=1\textwidth]{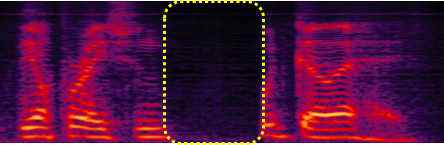}
         \caption{Audio-visual}
         \label{fig:five over x}
     \end{subfigure}
        \caption{Example spectra of overlapped, target clean,  audio-only separated, and audio-visual separated speech segments obtained using the \textit{TF masking} based speech separation front-end of Figure 1 (a). The spectrum portions circled using yellow dotted lines in (c) and (d) represent the  interfering speaker’s speech, which is almost removed in (d) after applying audio-visual speech separation.  }
        \label{fig:three graphs}
\end{figure}

\begin{table*}[htb!]
    \caption{Performance of different fine-tuning methods conducted on audio-visual multi-channel speech recognition systems. $\dagger$ and  $\ddagger$ denotes a statistically significant improvement is obtained over the pipelined CTC (sys.2) and LF-MMI (sys.6) systems.}
    \label{tab:my_label}
    \centering
    {\scalebox{0.95}{
    \begin{tabular}{c|ccc||c|cc||c|cc||c|cc}
    \toprule
    \multirow{3}{*}{Sys} &\multicolumn{3}{c||}{Fine-tuning} & \multicolumn{3}{c||}{TF masking} &\multicolumn{3}{c||}{Filter\&Sum} &\multicolumn{3}{c}{MVDR}\\
    \cline{2-13}
    &\multirow{2}{*}{Criterion} &\multirow{2}{*}{Sep.} &\multirow{2}{*}{Recg.} & Si-SNR &\multicolumn{2}{c||}{WER} & Si-SNR &\multicolumn{2}{c||}{WER} & Si-SNR &\multicolumn{2}{c}{WER} \\
    & & &                           &simu & simu &replay  & simu & simu&replay  & simu& simu&replay\\
    \hline
    \hline
    1& \multicolumn{3}{c||}{Not Applied}                           &10.16& 26.1 &28.0 &11.60& 23.5&30.9 &9.03& 26.1&25.8 \\
    \hline
    2& $\mathcal{L}_{CTC}$&{\xmark}     &{\cmark}                 &10.16& 22.9&23.2  &11.60& 19.2&24.1 &9.03& 19.3&17.3 \\
    3& $\mathcal{L}_{CTC}$&{\cmark}     &{\cmark}                 & 8.15& 19.3$^\dagger$&18.0$^\dagger$ &6.04& 17.2$^\dagger$&19.9$^\dagger$ &4.14& 18.6$^\dagger$&{{16.9}}$^\dagger$ \\
    4& $\mathcal{L}_{CTC}+\alpha\mathcal{L}_{Si-SNR}$&{\cmark}     &{\cmark}                &8.50& {\bf{18.6}}$^\dagger$&{\bf{18.0}}$^\dagger$ &9.17& {\bf{16.1}}$^\dagger$&{{\bf19.2}}$^\dagger$ &7.72& {\bf{18.4}}$^\dagger$&{{\bf16.9}}$^\dagger$ \\
    \hline
    \hline
    5& \multicolumn{3}{c||}{Not Applied}                           &10.16& 23.8 &26.2 &11.60& 21.1 &28.1 &9.03& 23.1 &22.8 \\
    \hline
    6& $\mathcal{L}_{LF-MMI}$&{\xmark}     &{\cmark}             &10.16& {{20.7}}&{{21.4}} &11.60& {{18.2}}&25.1 &9.03& 20.3&20.1    \\
    7& $\mathcal{L}_{LF-MMI}$&{\cmark}     &{\cmark}             &8.03&{{17.7}}$^\ddagger$&18.7$^\ddagger$ &9.20& {{16.9}}$^\ddagger$&22.4$^\ddagger$ &5.65& {{16.3}}$^\ddagger$&{{15.5}}$^\ddagger$        \\
    8& $\mathcal{L}_{LF-MMI}+\alpha\mathcal{L}_{Si-SNR}$&{\cmark}  &{\cmark}             &8.89& {\bf{17.7}}$^\ddagger$&\bf18.3$^\ddagger$&10.73& \bf16.6$^\ddagger$ &\bf21.6$^\ddagger$ &8.40& {\bf{16.1}}$^\ddagger$&{\bf{15.0}}$^\ddagger$        \\
    \bottomrule
    \end{tabular}}
    }
\end{table*}

\begin{table*}[htb!]
    \caption{Performance of different fine-tuning methods conducted on audio-visual multi-channel speech recognition systems. $\dagger$ and  $\ddagger$ denotes a statistically significant improvement is obtained over the pipelined CTC (sys.2) and LF-MMI (sys.6) systems.}
    \label{tab:my_label}
    \centering
    {\scalebox{1}{
    \begin{tabular}{c|ccc|c|cc|c|cc|c|cc}
    \toprule
    \multirow{3}{*}{Sys} &\multicolumn{3}{c|}{Fine-tuning} & \multicolumn{3}{c|}{TF masking} &\multicolumn{3}{c|}{Filter\&Sum} &\multicolumn{3}{c}{MVDR}\\
    \cline{2-13}
    &\multirow{2}{*}{Criterion} &\multirow{2}{*}{Sep.} &\multirow{2}{*}{Recg.} & Si-SNR &\multicolumn{2}{c|}{WER} & Si-SNR &\multicolumn{2}{c|}{WER} & Si-SNR &\multicolumn{2}{c}{WER} \\
    & & &                           &simu & simu &replay  & simu & simu&replay  & simu& simu&replay\\
    \hline
    \hline
    1& \multicolumn{3}{c|}{Not Applied}                           &10.16& 26.1 &28.0 &11.60& 23.5&30.9 &9.03& 26.1&25.8 \\
    \hline
    2& $\mathcal{L}_{CTC}$&{\xmark}     &{\cmark}                 &10.16& 22.9&23.2  &11.60& 19.2&24.1 &9.03& 19.3&17.3 \\
    3& $\mathcal{L}_{CTC}$&{\cmark}     &{\cmark}                 & 8.15& 19.3&18.0 &6.04& 17.2&19.9 &4.14& 18.6&{{16.9}} \\
    4& $\mathcal{L}_{CTC}+\alpha\mathcal{L}_{Si-SNR}$&{\cmark}     &{\cmark}                &8.50& {\bf{18.6}}&{\bf{18.0}} &9.17& {\bf{16.1}}&{{\bf19.2}} &7.72& {\bf{18.4}}&{{\bf16.9}} \\
    \bottomrule
    \end{tabular}}
    }
\end{table*}

\subsection{Performance of different fine-tuning methods}
The WER results of the audio-visual multi-channel system using different fine-tuning approaches aiming for integrating the separation front-end and recognition back-end are shown in Table II.
In these experiments, the visual features are used in both the separation and the  recognition components, while the AF features are adopted in the separation front-end only.
Before integration, both the separation and the recognition components are trained separately.
In the pipelined systems (sys.2, sys.5), the recognition back-ends are fine-tuned using the separation outputs, while the separation front-ends are kept unchanged.
In the jointly fine-tuned systems, both the separation and the  recognition components are fine-tuned using the recognition cost function (sys.2, sys.5) or multi-task criterion (sys.3, sys.6).
For the CTC based system, $\alpha$ is set as 0.1 for the \textit{TF masking} approach and 1 for the \textit{Filter\&Sum} and \textit{mask-based MVDR} approaches.
For the LF-MMI system, $\alpha$ is set as 0.01 (larger $\alpha$ will lead to performance degradation) for all the three channel integration approaches. 

\noindent Several trends can be observed from Table III:
\begin{enumerate}
    \item Compared with the audio-visual speech recognition systems without any separation front-ends in Table I (sys.10, sys.12), using the proposed audio-visual separation front-ends for these fixed recognition back-ends (sys.1, sys.5 in Table II)  reduce the WERs by up to 8.56\% and 6.09\% on the simulated and replayed data respectively.
    \item The jointly fine-tuned systems consistently outperform the pipelined systems for all the three channel integration methods (sys.2 vs. sys.3, sys.6 vs. sys.7), which confirms our arguments in section III-B.
    \item  Compared with the jointly fine-tuned systems using only the recognition cost, systems using multi-task criterion only provide marginal recognition improvements (sys.3 vs. sys.4, sys.7 vs. sys.8).
    \item Different from the trend in Table I, the LF-MMI based jointly fine-tuned systems  do not always outperform the CTC based systems, especially on the replay test set. 
    \item Jointly fine-tuning the separation and the recognition components using only the speech recognition cost degrades the speech separation performance in terms of Si-SNR (sys.2 vs. sys.3, sys.6 vs. sys.7) by up to 4.9dB. However, jointly fine-tuning these two components by multi-task criterion only degrades the Si-SNR performance by up to 1.27dB (sys.2 vs. sys.4, sys.6 vs. sys.8).
\end{enumerate}
Considering the average performance contrast between the CTC and LF-MMI costs fine-tuned systems over three beamforming methods on both the simulated and the replayed data in Table II (sys.3 vs. sys.6), we adopt jointly fine-tuned systems using only the CTC  cost function in all subsequent experiments.

\begin{table}[htb]
    \caption{Performance of audio-only and audio-visual overlapped speech recognition systems using various channel integration methods. The separation and the recognition components are jointly fine-tuned using the CTC loss. "AF" denotes angle feature.
    $\dagger$, $\ddagger$ and $^\star$ denotes a statistically significant improvement is obtained over the \textit{TF masking} (sys.5), \textit{Filter\&Sum} (sys.10) and \textit{mask-based MVDR} (sys.15) audio-only baseline systems.
    }
    \label{tab:my_label}
    \scalebox{0.85}{
    \centering
    {
    \begin{tabular}{c|c|c|c|c|c|c}
    \toprule
    \multirow{2}{*}{Sys}&\multicolumn{3}{c|}{Separation} & Recognition & \multicolumn{2}{c}{WER(\%)} \\
    \cline{2-7}
    & \multicolumn{1}{c|}{method}  & AF & +visual  &  +visual & simu & replay\\
    \hline
    \hline
    1&\multicolumn{3}{c|}{\multirow{2}{*}{raw channel 1}}              & \xmark & 75.36 & 80.55 \\
    2&\multicolumn{3}{c|}{}              & \cmark & 32.06 & 31.93\\
    \hline
    \hline
    3&\multirow{2}{*}{Delay\&Sum} & \cmark                         & - & \xmark & 49.25 & 44.34\\
    4& & \cmark                    & - & \cmark & 25.81 & 24.46\\
    \hline
    \hline
    5& \multirow{5}{*}{TF masking}    & \cmark                         & \xmark & \xmark & 33.12 & 46.75\\
    6&                                & \xmark                         & \cmark & \xmark & 24.64$^\dagger$ & 26.49$^\dagger$\\
    7&                                & \cmark                         & \cmark & \xmark & 23.17$^\dagger$ & 23.59$^\dagger$\\
    \cline{3-7}
    8&                               & \xmark                         & \cmark & \cmark & 21.32$^\dagger$ & 21.52$^\dagger$\\
    9&                                & \cmark                         & \cmark & \cmark & {\bf{19.25}}$^\dagger$ & {\bf18.03}$^\dagger$\\
    \hline
    \hline
    10&\multirow{5}{*}{Filter\&Sum}   & \cmark    & \xmark & \xmark & 30.24 & 43.83\\
    11&                                     & \xmark    & \cmark & \xmark &{23.09}$^\ddagger$ & 24.67$^\ddagger$\\
    12&                                     & \cmark   & \cmark & \xmark & {21.77}$^\ddagger$ & 24.66$^\ddagger$\\
    \cline{3-7}
    13&                                    & \xmark    & \cmark & \cmark & 21.02$^\ddagger$ & 20.02$^\ddagger$\\
    14&                                    & \cmark   & \cmark & \cmark & {\bf{17.21}}$^\ddagger$ & {\bf19.87}$^\ddagger$\\
    \hline
    \hline
    15&\multirow{5}{*}{Mask-based MVDR}   & \cmark                     & \xmark & \xmark & {25.38} & {39.07}\\
    16&                                   & \xmark                     & \cmark & \xmark & 23.96$^\star$ &  23.48$^\star$\\
    17&                                   & \cmark                     & \cmark & \xmark & 23.41$^\star$ & {21.17}$^\star$\\
    \cline{3-7}
    18&                                  & \xmark                     & \cmark & \cmark & {\bf17.34}$^\star$ &  {\bf16.21}$^\star$\\
    19&                                  & \cmark                     & \cmark & \cmark & 18.57$^\star$ & {16.85}$^\star$\\
    \bottomrule
    \end{tabular}}
    }
\end{table}

\begin{figure*}[bp]
\begin{minipage}{.5\textwidth}
\centering
\includegraphics[width=1\textwidth]{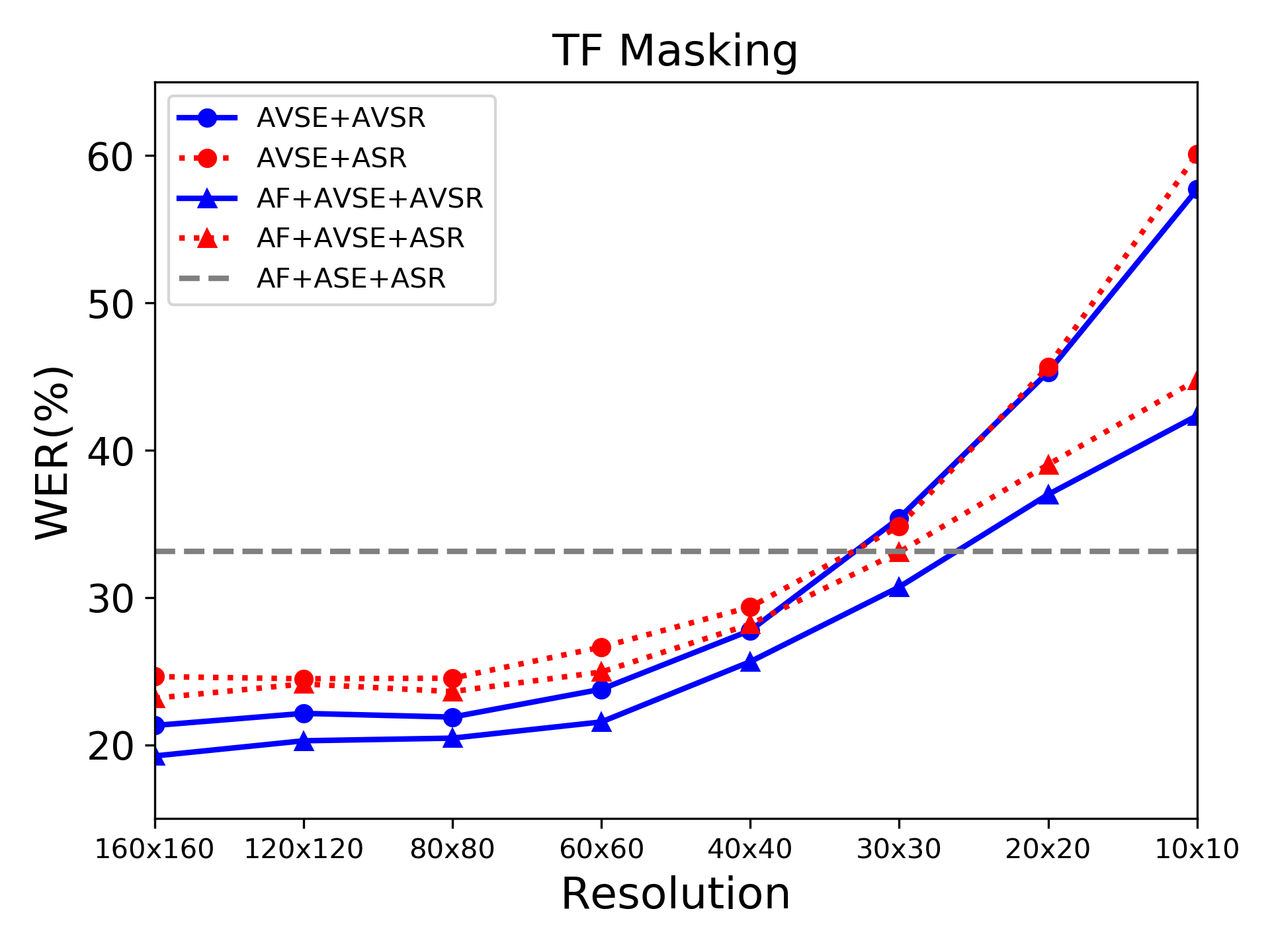}
\label{fig:fig1}            
\end{minipage}%
\begin{minipage}{.5\textwidth}
\centering
\includegraphics[width=1\textwidth]{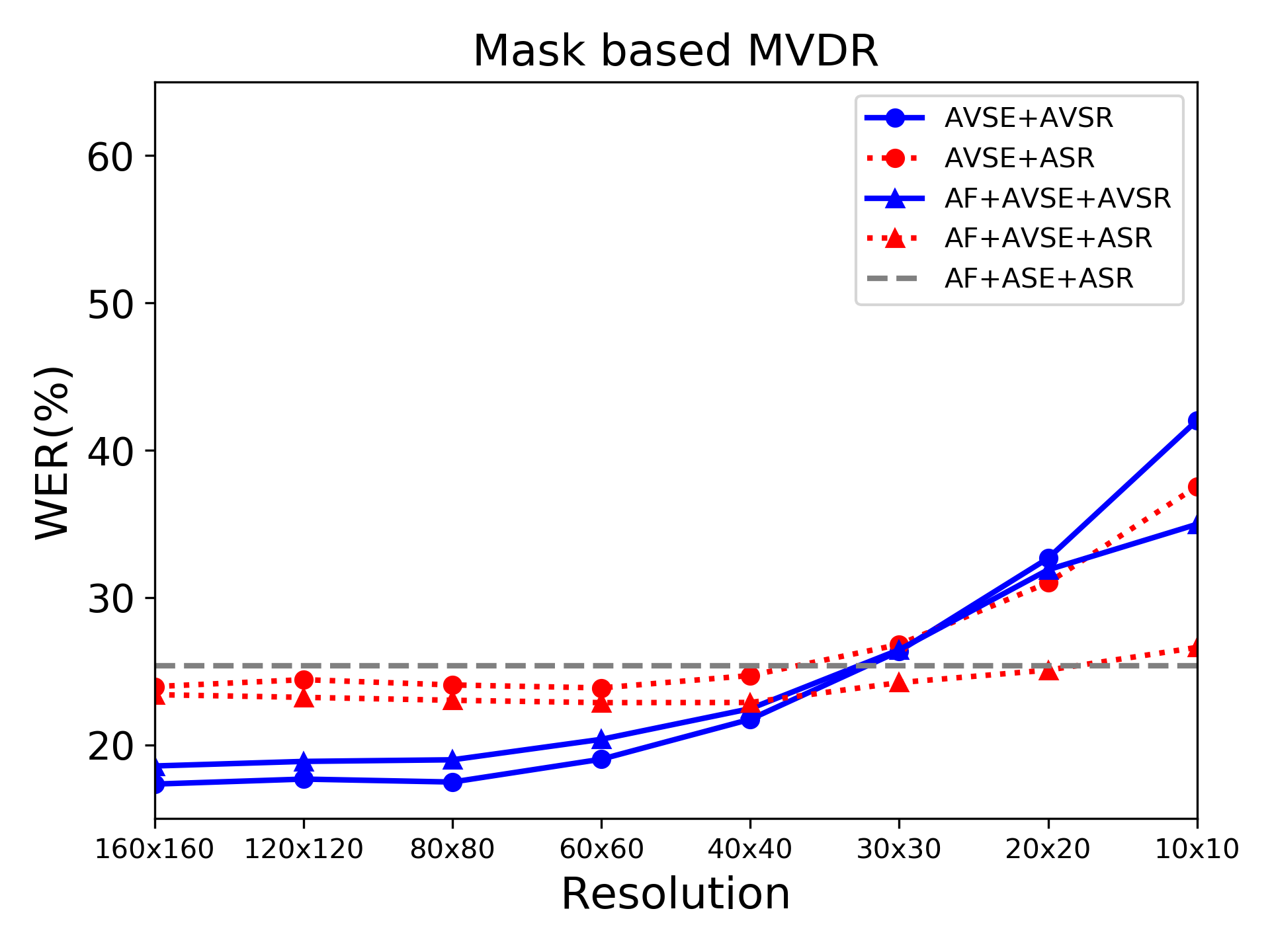}
\label{fig:fig2}            
\end{minipage}
\caption{
WER(\%) of \textit{TF masking} and \textit{mask-based MVDR} based AVSR systems of Table IV using different visual inputs resolutions ranging from $160\times160$ to $10\times10$.
"AVSE" and "AVSR" denote using visual modality in the separation front-end and recognition back-end respectively, "AF+" stands for optionally using angle features.
}
\end{figure*}
\subsection{Performance of audio-visual multi-channel AVSR systems}
In this section, we systematically investigate the performance improvements attributed to the visual modality in three types of audio-viusal multi-channel overlapped speech recognition systems featuring \textit{TF masking}, \textit{Filter\&Sum} and \textit{mask-based MVDR} neural beamformers. The visual modality’s impact on system performance is further analysed in a more advanced AVSR system configuration when it is used in combination with the angle features described previously in Section III-A.
To compare the performance between the conventional channel integration methods with the NN based methods, the traditional frequency domain delay and sum (\textit{Delay\&Sum}) beamformer is also adopted in this experiment.
The steering vectors used in such beamformer are computed based on the ground truth DOA for the simulated data and the approximated DOA for the replayed data. 

\noindent From Table IV, several trends can be observed:
\begin{enumerate}
    \item  Adding visual features can significantly improve the recognition performance on both the simulated and the replayed overlapped speech by up to 13.87\% and 28.72\% (sys.5 vs. sys.9), 13.03\% and 23.96\% (sys.10 vs. sys.14), 8.04\% and 22.86\% (sys.15 vs. sys.18) absolute WER reduction for the \textit{TF masking}, \textit{Filter\&Sum} and \textit{mask-based MVDR} approaches respectively. 
    \item When we only use the visual, but not the angle features in the proposed audio-visual multi-channel AVSR systems (sys.8, sys.13, sys.18), similar recognition performance is retained on both simulated and replayed data for all the three channel integration methods (sys.8 vs. sys.9, sys.13 vs. sys.14, sys.18 vs. sys.19).
    \item Using visual information in both the separation and the  recognition back-ends performs better than using visual information only in the separation front-ends. (sys.7 vs. sys.9, sys.12 vs. sys.14, sys.17 vs. sys.19)
    \item When we only use the angle features, a large performance gap between the simulated and the replayed data can be observed (sys.5,10,15). Since we use the ground truth DOA for the simulated data and approximated DOA for the replayed data, this phenomenon indicates that these three systems (sys.5,10,15) are sensitive to the precision of the DOA estimation. However, by adding visual features (sys.6-9, sys.11-14, sys.16-19), such performance gap is narrowed down greatly, which further confirms the efficacy of the visual modality.  
    \item The NN based separation front-ends (sys.5-19) outperform the conventional \textit{Delay\&Sum} beamformer (sys.3-4), which confirms the strength of the NN based channel integration methods. In addition, compared with the \textit{TF masking} (sys.5-9) and \textit{Filter\&Sum} (sys.10-14) approaches, the \textit{mask-based MVDR} systems (sys.15-19) show better performance on the replayed data set. 
\end{enumerate}

\subsection{Impact of low-resolution visual inputs}
In this section, we further investigate the robustness of the proposed \textit{TF masking} and \textit{mask-based MVDR} multi-channel AVSR systems in Table IV (sys.6-9 and sys.16-19) when lower resolution video inputs are used, as previously described in Section V-D.
Figure 8 shows the relationship between WER and visual input resolution for the \textit{TF masking} and the \textit{mask-based MVDR} based AVSR systems.
Several trends can be observed from Figure 8:
\begin{enumerate}
    \item  Although low-resolution visual inputs can cause performance degradation, the proposed systems consistently outperform the baseline audio-only systems even when the video resolution is aggressively reduced to as low as $40\times40$ pixels down from the full resolution of $160\times160$.
    \item The \textit{mask-based MVDR} ASR and AVSR systems are more robust to low resolution visual inputs than the \textit{TF masking} based comparable ASR and AVSR systems.
\end{enumerate}

\begin{table*}[htp]
    \caption{
    WER(\%) of CTC based \textit{TF masking} and \textit{mask-based MVDR} based AVSR systems when evaluated on data with visual occlusion ranging from 0\% up to 80\% coverage of the lip region.
    $\dagger$ and  $\ddagger$ denotes a statistically significant improvement is obtained over the \textit{TF masking} (sys.1) and \textit{mask-based MVDR} (sys.14) based audio-only systems.
    }
    \label{tab:my_label}
    \centering
    {
    \begin{tabular}{c|c|c|c|c|cc|c|ccccc}
    \toprule
    \multirow{2}{*}{Sys}&\multicolumn{3}{c|}{Separation}                & Recognition  &\multicolumn{2}{c|}{Training set} & \multirow{2}{*}{Test set}        &\multicolumn{5}{c}{WER(\%)} \\
    \cline{2-7}
    & \multicolumn{1}{c|}{method}  & AF & +visual  &  +visual      & no-occ & occ                              &  & 0\% & 20\% & 40\% & 60\% & 80\%  \\
    \hline
    \hline
    1&\multirow{13}{*}{TF masking}    & \cmark                         & \xmark & \xmark  &200h &\xmark &occ&\multicolumn{5}{c}{33.12}\\
    \cline{3-13}
    2&    & \xmark                         & \cmark & \xmark  &\multirow{4}{*}{200h} &\multirow{4}{*}{\xmark} &\multirow{4}{*}{occ}& 24.64$^\dagger$ & 30.02$^\dagger$ & 34.91 & 39.88 & 44.52 \\
    3&        & \cmark                     & \cmark & \xmark  &&&& 23.17$^\dagger$ &27.88$^\dagger$ & 32.48 &36.56 & 39.77 \\
    4&    & \xmark                         & \cmark & \cmark  &&&& 21.32$^\dagger$ & 28.30$^\dagger$  &34.86 &40.42 &44.83\\
    5&        & \cmark                     & \cmark & \cmark  &&&& \bf19.25$^\dagger$ & \bf24.82$^\dagger$ & \bf30.00$^\dagger$ &\bf33.39 &\bf37.90 \\
    \cline{3-13}
    6&    & \xmark                         & \cmark & \xmark  &\multirow{4}{*}{200h} &\multirow{4}{*}{\xmark} &\multirow{4}{*}{in-painting}& 24.64$^\dagger$& 31.06$^\dagger$ & 34.43 & 36.33 & 39.52 \\
    7&        & \cmark                     & \cmark & \xmark  &&&& 23.17$^\dagger$ &29.63$^\dagger$ & 31.13$^\dagger$ &33.84 & 35.22 \\
    8&    & \xmark                         & \cmark & \cmark  &&&& 21.32$^\dagger$ & 29.80$^\dagger$  &33.94 & 38.03 &41.91\\
    9&        & \cmark                     & \cmark & \cmark  &&&& \bf19.25$^\dagger$ & \bf26.22$^\dagger$ &\bf29.72$^\dagger$ &\bf32.17$^\dagger$ &\bf34.28 \\
    \cline{3-13}
    10&    & \xmark                         & \cmark & \xmark  &\multirow{4}{*}{200h} &\multirow{4}{*}{200h}  &\multirow{4}{*}{occ}& 24.43$^\dagger$ & 27.57$^\dagger$ & 29.8$^\dagger$ & 32.39 & 34.44 \\
    11&        & \cmark                     & \cmark & \xmark  &&&& 23.39$^\dagger$ &25.5$^\dagger$ &27.54$^\dagger$ &29.44$^\dagger$ &30.87$^\dagger$\\
    12&    & \xmark                         & \cmark & \cmark  &&&& 20.75$^\dagger$ & 24.46$^\dagger$  & 28.53$^\dagger$ & 32.76 & 35.12 \\
    13&        & \cmark                     & \cmark & \cmark  &&&& \bf18.57$^\dagger$ &\bf21.02$^\dagger$ &\bf23.78$^\dagger$ &\bf26.04$^\dagger$ & \bf28.21$^\dagger$ \\
    \hline
    \hline
    14&\multirow{13}{*}{Mask-based MVDR}    & \cmark    & \xmark & \xmark  &200h &\xmark&occ&\multicolumn{5}{c}{25.38}\\
    \cline{3-13}
    15&   & \xmark                     & \cmark & \xmark &\multirow{4}{*}{200h}&\multirow{4}{*}{\xmark} &\multirow{4}{*}{occ}& 23.96$^\ddagger$ & 24.55$^\ddagger$ & 26.70 & 28.59 & 31.20 \\
    16&  & \cmark                      & \cmark & \xmark &&&& 23.41$^\ddagger$ &23.81$^\ddagger$ &23.99$^\ddagger$ & 24.55$^\ddagger$ & 25.36\\
    17&   & \xmark                     & \cmark & \cmark &&&& \bf17.34$^\ddagger$ & \bf21.92$^\ddagger$ & \bf25.03 &29.62 &34.01 \\
    18&  & \cmark                      & \cmark & \cmark &&&& 18.57$^\ddagger$ &22.66$^\ddagger$ & 25.66 & \bf28.77& \bf31.61 \\
    \cline{3-13}
    19&   & \xmark                     & \cmark & \xmark &\multirow{4}{*}{200h}&\multirow{4}{*}{\xmark} &\multirow{4}{*}{in-painting} &23.96$^\ddagger$ & 25.03 & 25.50 & 26.79 & \bf27.35 \\
    20&  & \cmark                      & \cmark & \xmark &&&& 23.41$^\ddagger$ &24.28$^\ddagger$ &\bf24.47$^\ddagger$ & \bf24.41$^\ddagger$ &  \bf24.47$^\ddagger$\\
    21&   & \xmark                     & \cmark & \cmark &&&& \bf17.34$^\ddagger$ & \bf22.11$^\ddagger$ & 24.64$^\ddagger$ &27.72 &29.54 \\
    22&  & \cmark                      & \cmark & \cmark &&&& 18.57$^\ddagger$ &22.64$^\ddagger$ & 25.47 & 27.77& 29.13 \\
    \cline{3-13}
    23&   & \xmark                     & \cmark & \xmark &\multirow{4}{*}{200h}&\multirow{4}{*}{200h} &\multirow{4}{*}{occ}& 23.62$^\ddagger$ & 24.17$^\ddagger$ & 25.63 & 25.53 & 27.16 \\
    24&  & \cmark                      & \cmark & \xmark &&&& 22.85$^\ddagger$ & 23.35$^\ddagger$ & 23.62$^\ddagger$ & 24.16$^\ddagger$ & \bf24.27$^\ddagger$ \\
    25&   & \xmark                     & \cmark & \cmark &&&& \bf16.20$^\ddagger$ & \bf19.10$^\ddagger$ & \bf21.82$^\ddagger$ &23.75$^\ddagger$ & 26.74\\
    26&  & \cmark                      & \cmark & \cmark &&&& 18.08$^\ddagger$ &20.27$^\ddagger$ &22.27$^\ddagger$ & \bf23.60$^\ddagger$ &25.80\\
    \bottomrule
    \end{tabular}}
\end{table*}

\begin{figure*}[bp]
\begin{minipage}{.5\textwidth}
\centering
\includegraphics[width=1\textwidth]{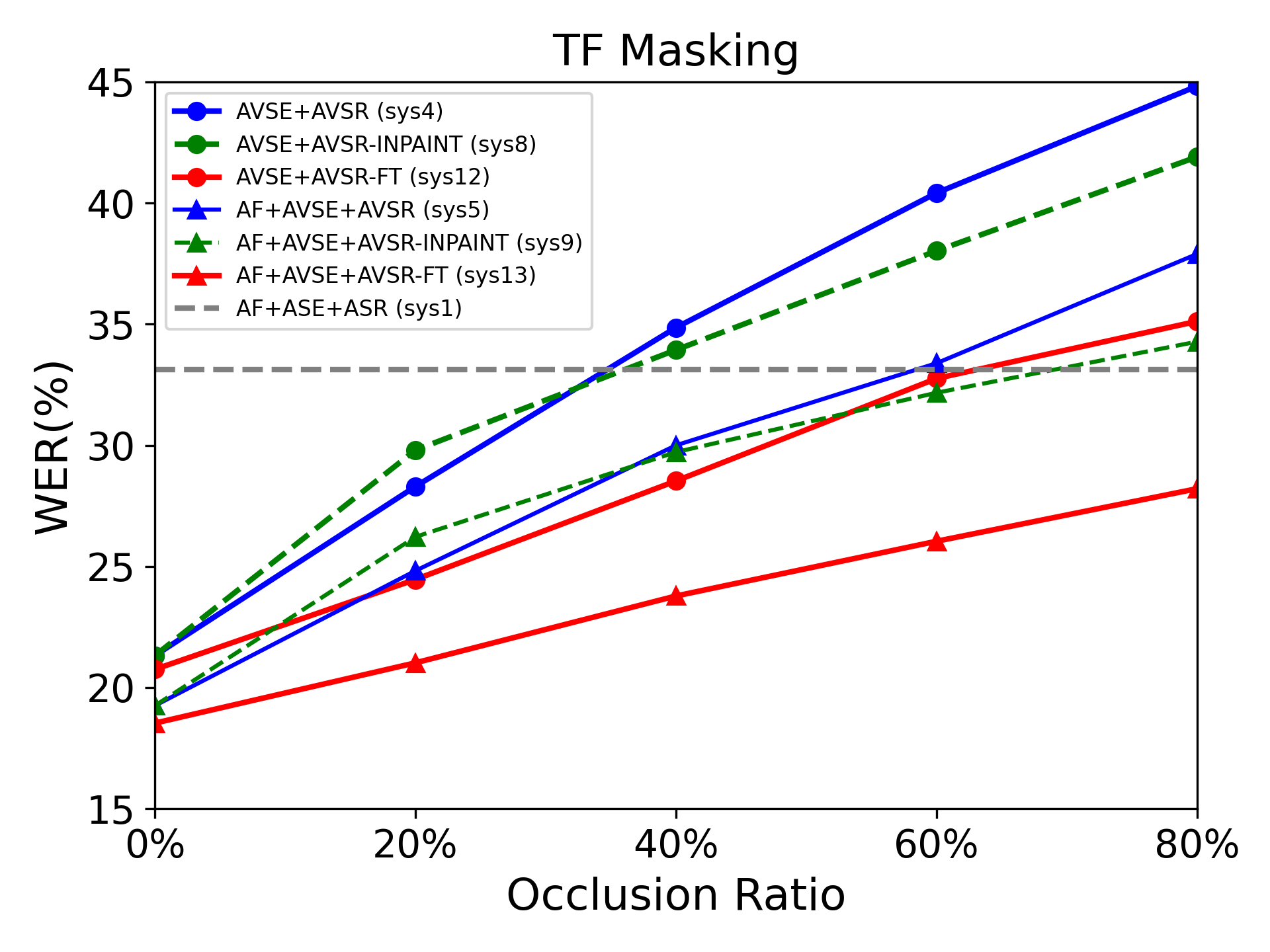}
\label{fig:fig1}            
\end{minipage}%
\begin{minipage}{.5\textwidth}
\centering
\includegraphics[width=1\textwidth]{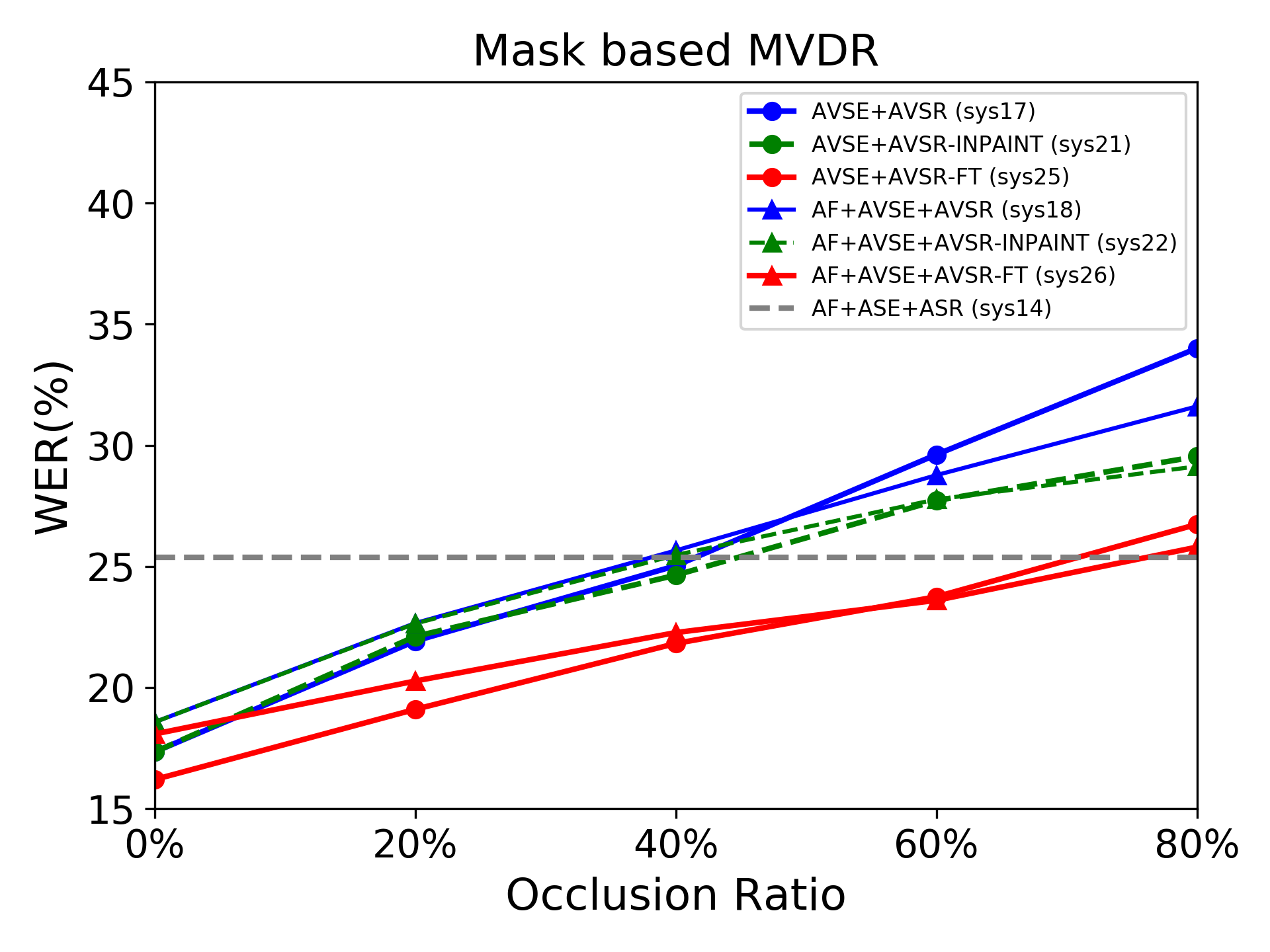}
\label{fig:fig2}            
\end{minipage}
\caption{
WER(\%) of the CTC based \textit{TF masking} and \textit{mask-based MVDR} based AVSR systems of Table V when evaluated on data with visual occlusion ranging from 0\% up to 80\% coverage of the lip region.
"AVSE" and "AVSR" denote using visual modality in the separation front-end and recognition back-end respectively, "AF+" stands for optionally using angle features, "+FT" denotes fine-tuning the system on multi-style data mixed with original and occluded video inputs,  "+INPAINT" denotes using the in-painting network to restore the occluded video.
}
\end{figure*}

\begin{figure}[htb]
    \centering
    \includegraphics[width=8.8cm]{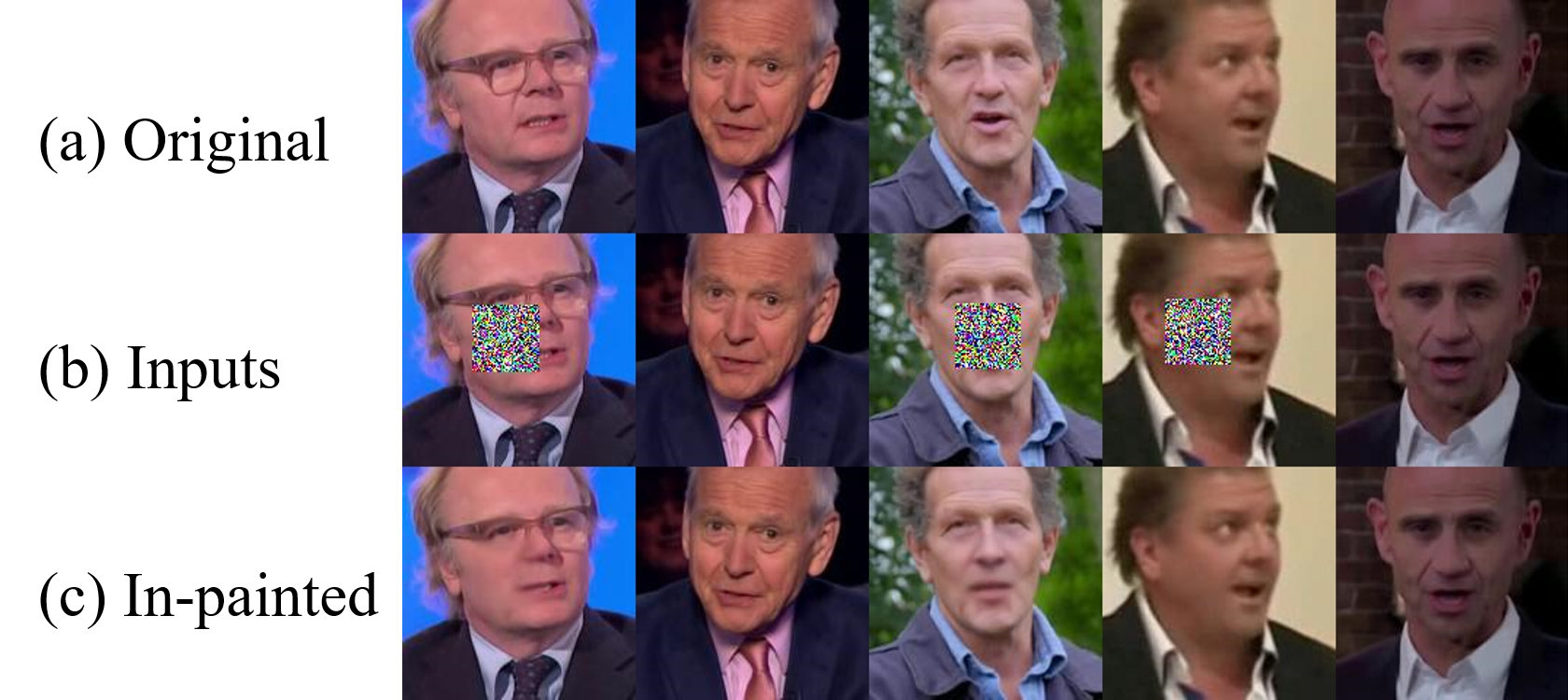}
    \caption{Examples of (a) original video snapshots; (b) randomly occluded video; and (c) restored video obtained using the in-painting network.}
    \label{fig:my_label}
\end{figure}

\subsection{Impact of visual occlusion}
In this section, we further investigate the robustness of the proposed AVSR multi-channel recognition systems when the video data quality is degraded by visual occlusion, as previously described in Section V-D.

As shown in Table V, three methods are considered to improve the system performance with occluded visual inputs:

\noindent 1) {\bf Using angle features}:
With the DOA information contained by angle features, the negative influence of visual input occlusion can be alleviated.
In the \textit{TF masking} systems, using AF consistently improves the system robustness to visual occlusions (sys.4 vs. sys.5, sys.8 vs. sys.9, sys.12 vs. sys.13). 
For the \textit{mask-based MVDR} systems, angle features show their effectiveness when the occlusion rate is larger than 60\% (sys.17 vs. sys.18, sys.25 vs. sys.26). 

\noindent 2) {\bf Using multi-style occluded data}: 
In order to improve the generalization to video occlusion at various percentage settings described in Section V-D, a 400-hour multi-style audio-visual training data set containing a 200h subset with video occlusion applied and the remaining half based on the original 200h data without occlusion was used to fine-tune the \textit{TF masking} and the \textit{mask-based} MVDR multi-channel AVSR systems in Table IV (sys.6-9 and sys.16-19) using the CTC cost function. 
A general trend can be found in Figure 9 (systems fine-tuned using multi-style occluded data shown on red lines) and Table V: the multi-style occluded data fine-tuned audio-visual multi-channel recognition systems (sys.13 and sys.26 in Table V) consistently outperform the baseline audio-only systems (Sys.1, 14 in Table V) even when using occluded video input with the lip region randomly covered up to 80\% for the TF masking (sys.1 vs. sys.13) and 60\% for the mask-based MVDR (sys.14 vs. sys.26) multi-channel systems.

\noindent 3) {\bf In-painting}: 
A visual in-painting neural network following \cite{inpainting} was trained using the occluded {\tt Train-val} set of LRS2 to in-paint the occluded visual image. Figure 10 shows some examples of the occluded images before and after being restored using the in-painting approach. If the input image is occluded, the in-painting network can restore the occluded region with some distortion. 
In contrast, if the video is not occluded, the in-painting network will keep the image almost unchanged. 
From Figure 9 (line in green vs. line in blue) and Table V (sys.2-5 vs. sys.6-9, sys.15-18 vs. sys.19-22) the following can be observed: using in-painting neural network can improve both the \textit{TF-masking} and the \textit{mask-based MVDR} multi-channel AVSR systems' robustness to visual occlusion when the occlusion rate is 60\% or above.


\noindent Several additional trends can be observed from Table V and Figure 9:
\begin{enumerate}
    \item As  shown  in  Figure  8,  the \textit{mask-based  MVDR} based systems are more robust to visual occlusion than the \textit{TF-masking} based systems. One possible explanation is that the \textit{mask-based MVDR} filter estimation exploits audio-video information across the entire speech segment and thus more robust to the partial, if not complete, occlusion being applied to the video data. This is different the other beamforming methods where no explicit constraint on using such longer span spatial-temporal contexts is enforced.
    \item The multi-style occluded data fine-tuning method outperforms the in-painting method (line in red vs. line in green, Figure 9). One possible explanation is that the in-painting network only use the visual information from the current occluded image frame to explicitly recover the occluded image, while during multi-style occluded data fine-tuning, both the speech separation front-end and the recognition back-end will learn the systematic variability among the occluded videos of the same audio contents but with different percentage of occlusion. This allows the resulting \textit{TF-masking} or \textit{mask-based MVDR} AVSR systems to implicitly build connection between the original and occluded data of the same audio and thus improves their robustness against video occlusion.
\end{enumerate}

\section{Conclusion}
In this work, we propose an audio-visual multi-channel  based  recognition  system  for  overlapped  speech. 
A series of audio-visual multi-channel speech separation front-ends based on \textit{TF masking}, \textit{Filter\&Sum}, and \textit{mask-based MVDR} are developed.
Jointly fine-tuning approaches are studied to integrate the separation and the  recognition components.
The impact of visual occlusion is also investigated. 
Experiments suggest that the proposed system significantly outperforms the baseline audio-only multi-channel ASR system on overlapped speech constructed using either simulation or replaying of the LRS2 dataset, which demonstrate the advantages of the visual information for overlapped speech recognition.
In the future, this work will be extended to: 
1) further integrating an audio-video de-reverberation component; 
2) multi-input multi-output AVSR systems facilitating speech separation and recognition for multi-talkers’ speech,
3) more advanced visual occlusion restoration methods to address visual occlusion issue.



%



\section*{Acknowledgment}
This research is supported by Hong Kong Research Grants Council GRF grant No. 14200218, 14200220, Theme-based Research Scheme T45-407/19N, Innovation \& Technology Fund grant No. ITS/254/19, and Shun Hing Institute of Advanced Engineering grant No. MMT-p1-19.

The authors would like to thank Yiwen Shao and Yiming Wang for the deep discussion about the LF-MMI implementation details.

\bibliographystyle{IEEEtran}
\bibliography{./mybib.bib}

\end{document}